\documentclass[%
 reprint,
jmp,
superscriptaddress,
 amsmath,amssymb,
 reprint,%
]{revtex4-2}

\usepackage{graphicx}
\usepackage{dcolumn}
\usepackage{bm}

\usepackage[utf8]{inputenc}
\usepackage[T1]{fontenc}
\usepackage{etoolbox}

\usepackage[english]{babel}
\usepackage{dcolumn}
\usepackage[version=3]{mhchem}
\usepackage{svg}
\usepackage{multirow}
\usepackage{comment}
\usepackage{tabularx}
\usepackage{xcolor}
\usepackage{array}
\usepackage{makecell}
\usepackage{amssymb}
\usepackage{float}
\usepackage{afterpage}
\usepackage[utf8]{inputenc}
\usepackage{braket}
\usepackage{amsmath}
\usepackage{amsfonts}
\usepackage{booktabs} 
\usepackage{titlesec}
\usepackage{dsfont}
\usepackage{pifont}
\usepackage{adjustbox}
\usepackage[normalem]{ulem}
\usepackage{tabularray}
\usepackage[left=0.8in,right=0.8in,top=0.8in,bottom=0.8in]{geometry}
\usepackage{upgreek}
\usepackage[super]{nth}
\usepackage{csquotes}

\newcommand{\revision}[1]{{\color{black}{#1}}}
\newcommand{\proof}[1]{{\color{black}{#1}}}

\date{\today}
             
\makeatletter
\def\@email#1#2{%
 \endgroup
 \patchcmd{\titleblock@produce}
  {\frontmatter@RRAPformat}
  {\frontmatter@RRAPformat{\produce@RRAP{*#1\href{mailto:#2}{#2}}}\frontmatter@RRAPformat}
  {}{}
}%
\makeatother
\begin{document}

\preprint{AIP/123-QED}

\title{Evaluation of the MACE Force Field Architecture: \\from Medicinal Chemistry to Materials Science}

\author{D\'{a}vid P\'{e}ter Kov\'{a}cs}
\affiliation{Engineering Laboratory, University of Cambridge, Cambridge, CB2 1PZ, UK}
 
 \author{Ilyes Batatia}
\affiliation{Engineering Laboratory, University of Cambridge, Cambridge, CB2 1PZ, UK}
\affiliation{ENS Paris-Saclay, Universit\'e Paris-Saclay, 91190 Gif-sur-Yvette, France}

\author{Eszter S\'{a}ra Arany}%
\affiliation{ School of Clinical Medicine, University of Cambridge, Cambridge, CB2 0SP}%

\author{G\'abor Cs\'anyi}
\affiliation{Engineering Laboratory, University of Cambridge, Cambridge, CB2 1PZ, UK}

\begin{abstract}
    The MACE architecture represents the state of the art in the field of machine learning force fields for \revision{a} variety of in-domain, extrapolation and low-data regime tasks. In this paper, we further evaluate MACE by fitting models for published benchmark datasets. We show that MACE generally outperforms alternatives for a wide range of systems from amorphous carbon, \proof{universal materials modelling,} and general small molecule organic chemistry to large molecules and liquid water. We demonstrate the capabilities of the model on tasks ranging from constrained geometry optimisation to molecular dynamics simulations and find excellent performance across all tested domains. We show that MACE is very data efficient, and can reproduce experimental molecular vibrational spectra when trained on as few as 50 randomly selected reference configurations. We further demonstrate that the strictly local atom-centered model is sufficient for such tasks even in the case of large molecules and weakly interacting molecular assemblies. 
\end{abstract}

\maketitle

\section{Introduction}
\label{sec:intro}

Machine learning force fields are becoming part of the standard toolbox of computational chemists as demonstrated by the increasing number of successful applications leading to new scientific discoveries, in fields including that of amorphous materials~\cite{deringer2021silicon_nature}, high-pressure systems~\cite{cheng2020highPhydrogen}, phase diagrams~\cite{kapil2022nano_water} and reaction dynamics of molecules~\cite{young2022reactionDynamics}.
These applications were enabled by a significant effort in developing a wide range of novel machine learning force field architectures. Recently, many of these were incorporated into a single, unifying design space~\cite{batatia2022designSpace, nigam2022unifiedCeriotti}, which helped uncover the relationship between seemingly dissimilar approaches such as the descriptor based machine learning force fields~\cite{behler2007generalized, bartok2013SOAP, shapeev2016moment, drautz2019atomic, nigam2020NICE, kovacs2021linear} and graph neural network based models~\cite{schutt2017schnet, gasteiger2020DimeNet++, schutt2021PAINN, batzner20223NequIP, tholke2022TorchMDNet}. This new understanding directly led to the MACE architecture~\cite{batatia2022mace}, which combines equivariant message passing with a high body-order description of the local atomic environment via spherical harmonic polynomials, leading to excellent performance in terms of accuracy, extrapolation, data and computational efficiency. 

In this paper, we evaluate the MACE architecture by training it on previously published datasets and benchmark its performance on a wide variety of tasks to demonstrate its out-of-the-box applicability across the chemical and physical sciences. The paper is organised as follows. Section~\ref{sec:theory} briefly reviews the MACE architecture and we also detail a training strategy that results in highly accurate models on all tested systems.  In Section~\ref{sec:locality}, we provide evidence that MACE can accurately describe the dynamics of large supra-molecular systems, including those critically controlled by intermolecular interactions. Section~\ref{sec:MACE_ANI} shows that MACE can fit a large, chemically diverse dataset of compounds made up of the H, C, N, and O chemical elements, improving on the previous state of the art by over a factor of three. In Section~\ref{sec:low_data}, we demonstrate that MACE can accurately fit the potential energy surface of small molecules from as little as 50 conformers. The resulting model is not only accurate on an independent test set but is also able to run long, stable molecular dynamics simulations without the need for any parameter tuning or any further iterative training. 
In Section~\ref{sec:carbon}, \ref{sec:hme21} and~\ref{sec:water}, we test MACE on condensed phase systems (carbon, disordered materials and liquid water), showing considerable improvements in accuracy compared to the models that were previously tested on these datasets. In the case of water, we also show that the resulting MACE model can accurately describe the thermodynamic and kinetic properties using NVT and NPT molecular dynamics simulations. 
Finally, in Section~\ref{sec:QM9} we evaluate the MACE architecture on the QM9 machine learning benchmark demonstrating that it also improves on the state of the art for several targets that are not force field fitting tasks. 

\section{The MACE architecture}
\label{sec:theory}
In this section we briefly review the MACE many-body equivariant message passing force field architecture, and then discuss a general training strategy relying on a loss scheduler. We also highlight the body-order of MACE models and discuss its implications for smoothness and extrapolation.

\subsection{Many-body equivariant message passing}
\label{sec:mace-detail}
The MACE model parametrises the mapping from the positions and chemical elements of the atoms to their potential energy. This is achieved by decomposing the total potential energy of the system into site energies (atomic contributions). The site energy of each atom depends on symmetric features that describe its chemical environment. MACE parametrises these features using a many-body expansion. To construct the features on the atoms first the local environment $\mathcal{N}(i)$ of atom $i$ has to be defined. $\mathcal{N}(i)$ is the set of all atoms $j$ in the system for which $|\bm{r}_{ij}| \leq r_{\text{cut}}$, where $\bm{r}_{ij}$ denotes the vector from atom $i$ to atom $j$ and $r_{\text{cut}}$ is a predefined cutoff (we discuss the effect of choosing different cutoff radii in Section~\ref{sec:locality}). Defining a local neighbourhood allows the construction of a graph from the geometry where the nodes are the atoms, and the edges connect atoms in each other's local environment. We denote the array of features of node (atom) $i$ by ${\bm h}_i$ and express them in the spherical harmonic basis and hence its elements are always indexed by $l$ and $m$.  The superscript on ${\bm h}^{(s)}$ indicates the iteration steps (corresponding to ``layers'' of message passing in the parlance of graph neural networks), and we denote the number of layers in the model by $S$. Equivariance of the model is achieved by utilising the transformation properties of spherical harmonics $Y_{l}^{m}$ under 3D rotations which are inherited by the node features with corresponding indices. 

We now describe the complete MACE model as it was introduced in Ref.~\cite{batatia2022mace}. In our discussion, we focus on showing both the key equivariant operations and also on identifying all free parameters of the model that are optimised during training. The equations are grouped together in a block for ease of reference. All free (``learnable'') parameters are denoted by the letter $W$, and all occurrences below correspond to {\em different} blocks of free parameters - the shapes of these parameter arrays are indicated by the use of different indices. Note, that in addition to the explicitly shown free parameters $W$, the fully connected multi-layer perceptrons (MLPs) also contain internally further learnable parameters. 

We start by initialising the node features ${\bm h}^{(0)}_{i}$ as a (learnable) embedding of the chemical elements with atomic number $z_i$ into $k$ learnable channels, c.f. Eq.~\eqref{eq:element_embedding}. This kind of mapping has been used extensively for graph neural networks~\cite{schutt2017schnet, schutt2021PAINN, batzner20223NequIP, gasteiger2020DimeNet++} and elsewhere~\cite{willatt2018feature, gubaev2019accelerating} and has been shown to lead to some transferability between molecules with different elements~\cite{batatia2022designSpace}. The zeros for the $lm$ indices correspond to these initial features being scalars, i.e. rotationally invariant. The higher order elements of ${\bm h}^{(0)}_{i}$ with nonzero $lm$ indices are implicitly initialised to zero.
 At the beginning of each subsequent iteration, the node features (both scalars and higher order) are linearly mixed together resulting in ${\bm{\bar h}}_{j}$, c.f. Eq.~\eqref{eq:linear_first}.

Next, we combine the features of each of the neighbouring atoms (edge in the graph), with the interatomic displacement vectors pointing to them from the central atom expressed using radial and spherical harmonics basis. This is analogous to the construction of the one-particle basis of neighbour density representations, such as SOAP\cite{bartok2013SOAP} and ACE\cite{drautz2019atomic,batatia2022designSpace} and we construct it in a similar way to Cormorant~\cite{anderson2019cormorant} and NequIP~\cite{batzner20223NequIP}. The relationships between these different approaches are discussed in Ref.~\cite{batatia2022designSpace}. We construct the radial basis set using the first spherical Bessel function $j^n_0$ for different wavenumbers, $n$, up to some small maximum (typically 8), in Eq.~\eqref{eq:rad_feats} as proposed in Ref.~\cite{gasteiger2020DimeNet++}. The Bessel functions are multiplied by a polynomial cutoff function $f_{\text{cut}}(r_{ij})$ which goes to zero smoothly at $r_{ij} = r_{\text{cut}}$. The radial information is then passed through an MLP, Eq.~\eqref{eq:radial_MLP}, whose inputs are the Bessel functions with different frequencies, and have many outputs, indexed by $(\eta_1,l_1,l_2,l_3)$, whose purpose is explained next. When combining positional information (itself an equivariant that transforms under rotation like a vector) with equivariant node features, we use the spherical tensor product formalism of angular momentum addition~\cite{wigner2012group}. All possible combination of equivariants are constructed using the appropriate Clebsch-Gordan coefficients, Eq.~\eqref{eq:phi-basis-t}. 
\revision{This operation is currently implemented using \texttt{e3nn}~\cite{geiger2022e3nn}}. There are multiple ways of constructing an equivariant combination with a given symmetry corresponding to $(l_3,m_3)$, and these multiplicities are enumerated by the index $\eta_1$. The large number of outputs (and their associated learnable weights) from the radial MLP are used as separate degrees of freedom. 

\begingroup\makeatletter\def\f@size{8.5}\check@mathfonts
\def\maketag@@@#1{\hbox{\m@th\large\normalfont#1}}%
\begin{align}
\label{eq:element_embedding}
h_{i,k00}^{(0)} &= \revision{\sum_z} W_{kz} \delta_{zz_{i}}\\
\vbox to 20pt{}
\label{eq:linear_first}
    \bar{h}^{(s)}_{i,kl_2m_2} &= \sum_{\tilde{k}} W_{k\tilde{k}l_2}^{(s)} h^{(s)}_{i,\tilde{k}l_2m_2} \\
\label{eq:rad_feats}
j^{n}_{0} (r_{ij}) &=  \sqrt{\frac{2}{r_{\text{cut}}}} \frac{\sin{\left(n\pi\frac{r_{ij}}{r_{\text{cut}}} \right)}}{r_{ij}} f_{\text{cut}}(r_{ij}) \\
\vbox to 20pt{}
\label{eq:radial_MLP}
    R_{k \eta_{1} l_{1}l_{2} l_{3}}^{(s)}(r_{ij}) &=   {\rm MLP}\left( \left\{ {j_0^n} (r_{ij})\right\}_{n}\right) \\
\vbox to 20pt{}
  \label{eq:phi-basis-t}
  \phi_{ij,k \eta_{1} l_{3}m_{3}}^{(s)} &= 
    \sum_{l_1l_2m_1m_2} C_{\eta_1,l_1m_1l_2m_2}^{l_3m_3}
      R_{k \eta_{1} l_{1}l_{2} l_{3}}^{(s)}(r_{ij}) \,\,\times \notag\\
      & \qquad\qquad \times Y^{m_{1}}_{l_{1}} (\boldsymbol{\hat{r}}_{ij}) \bar{h}^{(s)}_{j,kl_2m_2} \\ 
\vbox to 20pt{}
\label{eq:atomic-basis-t}
    A_{i,kl_{3}m_{3}}^{(s)} &= \sum_{\tilde{k}, \eta_{1}} W_{k \tilde{k} \eta_{1}l_{3}}^{(s)}
    \sum_{j \in \mathcal{N}(i)}  \phi_{ij,\tilde{k} \eta_{1} l_{3}m_{3}}^{(s)} \\
\vbox to 20pt{}
\label{eq:product-basis}
{\bm A}^{(s),\nu}_{i,k\bm l \bm m} &= \prod_{\xi = 1}^{\nu} A_{i,k l_\xi  m_\xi}^{(s)}\\
\vbox to 20pt{}
\label{eq:symmbasis_L1}
  {\bm B}^{(s),\nu}_{i,\eta_{\nu} k LM}
  &= \sum_{{\bm l}{\bm m}} \mathcal{C}^{LM}_{\eta_{\nu} \bm l \bm m} {\bm A}^{(s),\nu}_{i,k\bm l \bm m} \\
 \vbox to 20pt{}
 \label{eq:message}
  m_{i,k LM}^{(s)} &=  \sum_{\nu}\sum_{\eta_{\nu}} W_{z_{i} \eta_{\nu} k L}^{(s),\nu} {\bm B}^{(s),\nu}_{i,\eta_{\nu} k LM}\\
 \vbox to 20pt{}
 \label{eq:update}
  h^{(s+1)}_{i,k LM}
    &= \sum_{\tilde{k}} W_{k L,\tilde{k}}^{(s)} m_{i,\tilde{k}LM}^{(s)}
  + \sum_{\tilde{k}} W_{kz_{i} L,\tilde{k}}^{(s)}
  h^{(s)}_{i,\tilde{k}LM} 
\end{align}\endgroup

The one-particle basis $\phi$ is summed over the neighborhood in Eq.~\eqref{eq:atomic-basis-t}, this is where permutation invariance of the MACE descriptors over the atoms in the neighbourhood is achieved - note that the identity of chemical elements has already been embedded, and hence this sum is over all atoms, regardless of their atomic number. A linear mixing of $k$ channels with learnable weights yields the atomic basis, $A_i$ (c.f. ACE~\cite{drautz2019atomic, batatia2022designSpace}). 

In an analogous fashion to the ACE construction, many-body symmetric features are formed on each atom by taking the tensor product of the atomic basis, $A$, with itself $\nu$ times, yielding the ``product basis'', ${\bm A}_i$ (c.f. Eq.~\eqref{eq:product-basis}). Note that in forming the tensor product, each $k$ channel is treated independently. This method of {\em tensor sketching} has been widely used in signal processing~\cite{sidiropoulos2017tensor} and was formally shown to not degrade the expressibility of many-body models while substantially reducing computational cost compared to a full tensor product~\cite{darby2022TrACE}. \revision{The tensor product of Equation~\eqref{eq:product-basis} is implemented using an efficient loop tensor contraction algorithm~\cite{batatia2022mace}. }

In Eq~\eqref{eq:symmbasis_L1} the product basis is contracted to yield the fully symmetric basis, $\bm B_i$, using the generalized Clebsch-Gordan coefficients, $\mathcal{C}$, where again there are multiple ways to arrive at a given output symmetry and these are enumerated by $\eta_{\nu}$. The bold ${\bm l \bm m}$ signify that these are multi-indices (an array of indices), and the bold styles of $\bm A$ and $\bm B$ is a reminder that these are many-body features. The maximum body-order is controlled by limiting $\nu$. 

Finally, a ``message'' $m_{i}$ is formed on each atom as a learnable linear combination of the symmetrized many-body features of the neighbours as shown in Eq.~\eqref{eq:message}. To form the node features of the next layer we are adding this message to the atoms' (nodes') features from the previous iteration using weights that depend explicitly on the atoms' chemical element ($z_i$), as shown in Eq.~\eqref{eq:update}. 
Note, that because the initial node features ${\bm h}^{(0)}$ are solely functions of the chemical element corresponding to the node, in the first layer the second term of Eq.~\eqref{eq:update} is omitted. This allows setting and fixing the energy of isolated atoms (i.e. those with no neighbours in their environment)~\cite{batatia2022designSpace}, which is often desirable~\cite{deringer2021gaussian}.

Equations~\eqref{eq:linear_first}-\eqref{eq:update} comprise a MACE layer, and multiple layers are built by iteration. This means that the effective receptive field of the model (the region around an atom from which information is used to determine the site energy of the atom) is approximately a sphere of radius $S \times r_{\text{cut}}$. More precisely, a neighbouring atom contributes to the site energy if it can be reached in $S$ hops on the graph defined above. The selected clusters form a sparse set of all possible clusters in $S \times r_{\text{cut}}$. Therefore message passing methods can be viewed as sparsifications of larger atom-centered methods \cite{batatia2022designSpace}. In practice, we almost always use two layers, so $S=2$. 

The output of the model, the site energy, is a learnable combination of the rotationally invariant part of the node features, 
\begin{equation}
  \label{eq:site-energy}
    E_i = \sum_{s=1}^S E_i^{(s)} = \sum_{s=1}^S
    \mathcal{R}^{(s)} \left({\bm h}_{i}^{(s)}\right),
\end{equation}
where $\mathcal{R}$ is a linear map for the first layer features and a shallow one hidden layer MLP for the second layer features,
\begin{equation}\label{eq:readout}
    \mathcal{R}^{(s)} \left( \boldsymbol{h}_i^{(s)} \right) = 
    \begin{cases}
      \sum_{k}W^{(s)}_{k}h^{(s)}_{i,k00}     & \text{if} \;\; s < S \\[13pt]
      {\rm MLP} \left( \left\{ h^{(s)}_{i,k00} \right\}_k \right)  &\text{if} \;\; s = S
    \end{cases}
\end{equation}
Keeping the readout function linear for all but the last layer helps preserve the body-ordered nature of the model, see Sec.~\ref{sec:body-order} and Ref.~\cite{batatia2022designSpace}. The forces on the atoms are determined as usual by taking analytical derivatives of the total potential energy, using auto-differentiation tools,
\begin{equation}
  \label{eq:forces}
       \bm{F} = -\nabla \sum_{i} E_{i}
\end{equation}

Neural network models such as MACE have a very large number of free parameters, especially compared with linear or kernel based models typically applied to the same tasks. It is thought that this overparameterisation helps the training when using stochastic gradient descent and brings accuracy and regularity~\cite{geiger2020scaling, Jiang2020Fantastic}. The trade-off between the size (and therefore computational cost) of the model and its accuracy needs to be controllable. The key model size controls in MACE are the number of embedding channels $k$ and the highest order $L_{\text{max}}$, of the symmetric features ${\bm B}^{(s),\nu}_{i,\eta_\nu k LM}$. In this paper, we will refer to the different sized models using these two numbers; for example, the invariant MACE model (corresponding to $L _{\text{max}}=0$) with 64 channels is denoted \texttt{MACE 64-0}. 

\revision{The computational cost of running a MACE model depends on the details of the systems studied (e.g. the average number of neighbours) and the hyperparameters that control the size of the model. The latency of the models (i.e. the shortest time to calculate forces on small systems, excluding calculation of the neighbour list and other book-keeping necessary for running MD) ranges from 
0.7-20 ms, corresponding to small (\texttt{MACE 64-0}) or larger (\texttt{MACE 256-2}) models as measured on an Nvidia A100 GPU. A comprehensive assessment of MD performance, which we leave to future work, needs to consider the relationship between model size, accuracy and execution speed. It is also useful to report the {\em fastest possible} MD performance, which we take to be that of the smallest model that can run stable MD for an arbitrary (in practice very large) number of steps. At present the performance of MACE in MD simulations is between 2-50M steps / day (a million steps corresponds to a nanosecond with 1 femtosecond time step), depending on model size and the details of the full software stack, with the highest performance achieved using the JAX version of MACE using JAX-MD \cite{jaxmd2020}, see e.g. the salicylic acid example ran using the \texttt{64-0} model. The evaluation\proof{\sout{s}} speeds in LAMMPS and OpenMM are currently somewhat slower, but new custom CUDA kernels and improved MD interfaces are actively being developed that will likely enable even higher performance. The training times of the MACE models also vary vastly depending on the dataset size and the model size, and typically ranges from 1 hour to 1 day on a single A100 GPU for the models shown here. 

The computational performance of MACE compares very favourably to other alternative equivariant neural network potentials. This is primarily the result of MACE forming the high body-ordered features very efficiently via the symmetric tensor products carried out on the nodes. For example the Allegro model, which is one of the leading architectures, that uses tensor products between the features on the edges of the graph rather than nodes, has a fastest reported speed of 18M steps per day~\cite{musaelian2023Allegro}. It should be noted though, that the Allegro model has already been demonstrated to scale well to many millions of atoms \proof{on many GPUs}~\cite{musaelian2023Allegro_scaling}. This is in principle possible for MACE, which is also a strictly local model with a similarly moderate size receptive field, 5-10~\AA, though the necessary LAMMPS interface implementation is ongoing work and will be demonstrated in future publications. Allegro typically uses 4-6~\AA\ for production runs and larger for accuracy benchmarks.}

\subsection{The body-order of MACE models}
\label{sec:body-order}
It is interesting to consider the node features ${\bm h}_i$ of MACE as body-ordered descriptors which, in the limit of complete basis of radial functions and spherical harmonics, fully linearly span the space of symmetric functions over chain like clusters with hops of size $r_{cut}$, up to the maximum body-order of the features~\cite{dusson2022atomic, batatia2022designSpace, darby2022TrACE}. In the MACE models studied in this paper, each layer has a body-order of 4 (corresponding to $\nu=3$ in Eq,~\eqref{eq:product-basis}). Therefore, the node features of the first layer are 4-body functions. They are then expanded in the second layer in the second one-particle basis, which are $4+1=5$-body functions, since each application of the one-particle basis adds one to the body-order on account of the central atom. Then, taking the tensor products of these features with themselves three times, we obtain $3\times4+1=13$-body features (note that the 5-body terms of the first layer all share the same central atom). This mechanism of efficiently forming very high body-order linearly complete features is unique to the MACE architecture and it might be one of the reasons underpinning its excellent performance in the low-data regime and extrapolation.

\subsection{Loss scheduler}

We found that to get the best performance from the MACE model, particular care has to be taken with the weight factors in the loss function. The training loss, $\mathcal{L}$,  is typically computed as the weighted sum of the mean squared errors of the total energy, the force components (and virials or stresses if available for periodic systems), e.g. 
\begingroup\makeatletter\def\f@size{8}\check@mathfonts
\def\maketag@@@#1{\hbox{\m@th\large\normalfont#1}}%
\begin{align}
  \label{eq:loss_fn}
  \mathcal{L} = 
  \frac{\lambda_{E}}{B} \sum_{b=1}^{B} &\left(\frac{E_{b} - \hat{E}_{b}}{N_{b}} \right)^{2} + \\ \notag
  &+ \frac{\lambda_{F}}{3B} \sum_{b=1}^{B} 
  \sum_{i_{b},\alpha=1}^{N_{b}, 3} 
  \left(- \frac{\partial E_{b}}{\partial r_{i_{b},\alpha}} - \hat{F}_{i_{b},\alpha} \right)^{2} ,
\end{align}
\endgroup
where the sum is taken over the atomic configurations $b$ in the current batch with batch size $B$, $E_b$ is the model's prediction of the total energy, $\hat{E}_b$ and $\hat{F}_b$ are the training data corresponding to total energy and force, respectively. The number of atoms in the configuration is $N_b$, atoms are indexed by $i_b$ with elements of their position vector denoted by $r_{i_b,\alpha}$, in the Cartesian direction $\alpha$. The weights of energies and forces are controlled by $\lambda_E$ and $\lambda_F$. The choice of these weights is crucial, and we observed that the most accurate models are obtained when we train with a high weight on the forces compared to the other properties. A possible reason for this could be that the forces are local quantities and contain information about the dependence of the total energy on each atom. However, having very accurate force predictions does not necessarily result in accurate energy predictions. This is especially the case in systems where the training set is heterogeneous, by which we mean that it is composed of a wide variety of different systems, well separated in atomic configuration space, e.g. different molecules, or phases of a solid with very different structure. In this case, the model can accurately learn the local potential energy surface of each system individually, but it might not learn their relative energies correctly, resulting in large absolute energy errors. 

To reduce absolute energy errors, we use a loss scheduler. For about $60\%$ of the total training time, $\lambda_{F} > \lambda_{E}$ is used. For the second part of the training, the weights are switched so that $\lambda_{E} > \lambda_{F}$, and the learning rate is decreased by a factor of 10. In this way, the absolute energy errors can be reduced while keeping the force predictions' high accuracy. An example of the energy validation error dramatically improving as the loss is changed is given in the Appendix Figure~\ref{fig:loss_schedule}. \revision{The Figure also compares the two phase training with using the weights of the second phase from the beginning and shows that the two phase protocol not only accelerates the convergence, but also results in overall decreased errors. } Note, that many of the experiments in the paper were carried out using an earlier version of the loss function which also too the average over the force components, rather than the sum as shown in Appendix~\ref{sec:old_loss}. 

\section{Locality of large molecular systems}
\label{sec:locality}

In this section, we examine the locality assumption used by MACE and compare its performance against a global machine learning force field, sGDML, which does not use a cutoff to  decompose the energy into local site energy terms~\cite{chmiela2018towards}. \proof{In particular, we compare MACE to the recent improved version of sGDML which uses a reduced set of global descriptors to allow for the fitting of systems with hundreds of atoms~\cite{kabylda2023efficient_sGDML}. Furthermore, we also compare MACE to the VisNet-LSRM model~\cite{li2023LongShortMP} which employs a mixed short-range long-range description of the potential energy surface by combining local message passing with message passing between larger fragments.} 

As explained in Sec.~\ref{sec:mace-detail} the effective cutoff in MACE is the number of layers times the cutoff distance in each layer. We also examine to what extent the information transfer between layers is effective and how property predictions of large molecular systems differ between a two-layer and a single-layer MACE model with the same receptive fields. 

For this study, we use the recently published MD22 dataset, which was designed to be challenging for short-range models~\cite{MD22}. The datasets include large molecules and molecular assemblies containing hundreds of atoms with complex intermolecular interactions. The dataset was created by running {\em ab initio} molecular dynamics simulations at elevated temperatures to better sample the configuration space of the systems. The training set size was determined such that the total energy error of the sGDML model is below chemical accuracy,  1 kcal/mol. Our study uses the same training set sizes for a fair comparison.  

\begin{table}[H]
\centering
  \caption{
    \textbf{Global vs Long-range vs Local models on MD22 dataset of large molecules.}
    Energy (E, meV/atom) and force (F, meV/\AA) mean absolute errors (MAE) of models. The approximate diameter of the system is denoted by $d$.
  }
\resizebox{0.51\textwidth}{!}
     {%
\begin{tabular}{m{2.5cm} c c c c c c c} 
  \toprule
     &  \qquad $d$~(\AA) \qquad& &  \textbf{MACE}  & \textbf{MACE}    & \textbf{MACE} & \textbf{VisNet-LSRM} &\textbf{sGDML} \\
     &    &   &  256-2  &  256-0   &   256-2  &   &  \\
   Cutoff distance  &  &   & $2\times3$\AA   & $1\times6$\AA   &  $2\times5$\AA & Long-range & Global  \\
  \midrule
  \multirow{2}{*}{Tetrapeptide} &
  \multirow{2}{*}{ $\sim$12}
                & E  & 0.608  &  0.345  &  \textbf{0.064} & 0.080 & 0.40   \\
              &  & F  & 7.6  &  17.0 &  \textbf{3.8} & 5.7 & 34    \\ \hline
  \multirow{2}{*}{Fatty acid} &
  \multirow{2}{*}{ $\sim$16}
                & E  & 0.446 & 0.399 & 0.102 &  \textbf{0.058} & 1.0 \\
              &  & F &  6.2 & 23.5 & \textbf{2.8}  & 3.6  & 33 \\ \hline
  \multirow{2}{*}{Tetrasaccharide} &
    \multirow{2}{*}{ $\sim$14}
                & E  & 0.252  & 0.357  & 0.062  & \textbf{0.044}  & 2.0 \\
             &   & F &  6.8 &  27.0 & \textbf{3.8} &  5.0 &  29 \\ \hline \\[1pt]
  Nucleic acid &
    \multirow{2}{*}{ $\sim$22}
                & E  &   0.902  &   0.155  & 0.079  &  \textbf{0.055} & 0.52  \\
        (AT-AT)     &   & F  &  13.3  &  14.9  &  \textbf{4.3} &  5.2  &  30  \\[1pt] \hline \\[1pt]
  Nucleic acid &
    \multirow{2}{*}{ $\sim$24}
                & E &  0.603  &  0.166   & 0.058  & \textbf{0.049}  & 0.52 \\
      (AT-AT-CG-CG)  &   & F &  16.3 &  20.1  & \textbf{5.0}  &  8.3  & 31  \\ \hline
  \multirow{2}{*}{\text{Buckyball catcher}} &
    \multirow{2}{*}{ $\sim$15}
                & E &  0.476  &  0.171 &  0.141 & \textbf{0.124}  &  0.34  \\
            &    & F  &  13.1 &  22.2  &  \textbf{3.7} & 11.6  &  29  \\ \hline 
  Double-walled &
    \multirow{2}{*}{ $\sim$33}
                & E &   0.207  & 0.231 &  0.194 & \textbf{0.117}  &  0.47 \\
      nanotube       &   & F &  17.9  & 39.6 &   \textbf{12.0}  &  28.7  &  23   \\
  \bottomrule
\end{tabular}

}
  \label{tab:md22}
\end{table}
\subsection{Effect of locality on energy and force errors}

In order to test the influence of the receptive field of local models on their accuracy, we trained a series of three MACE models with different combinations of cutoff distances and number of layers. The typical MACE model with two layers and $5$~\AA\ cutoff at each layer is compared to a one layer MACE with $6$~\AA\ cutoff and a two layer MACE with both layers having $3$~\AA\ cutoffs.

The best performing MACE model, which employs a $2 \times 5$ ~\AA\ cutoff, improves the errors of the sGDML model by up to a factor of 10. Crucially, even this 10~\AA\ receptive field is considerably smaller than the diameter of the systems in the dataset, so the MACE model is effectively local. Even the $2 \times 3$~\AA\  and $1 \times 6$~\AA\ models outperform sGDML for all systems. This is likely because the strength of local intramolecular (covalent) interactions is much higher than that of intermolecular interactions. So better overall accuracy can be achieved by only improving the short-range description, which MACE evidently does. In the Appendix, we show the vibrational spectrum of the tetrapeptide, showing excellent agreement between all three MACE models. \revision{When comparing MACE to VisNet-LSRM, the best model reported to date, we find that the strictly local MACE model has significantly lower force errors but in most cases the long-range model has lower energy errors, though it should be noted that the energy errors of both models are in the order of 0.1 meV / atom or lower in most cases. This small, but consistent improvement in the energies might result from more accurate description of remaining small long-range interactions beyond the 10~\AA\ cutoff of our longest range MACE model, or from different relative force and energy weights used during training. }  It is not trivial to come up with benchmarks that specifically test the description of long range interactions and might pose a difficulty for short range models. 

Typical intermolecular interactions appear to be well captured at the 5-6~\AA\ range. The evidence for this is in the energy errors for the nucleic acid and the Bucky-ball catcher systems. In these systems, the intermolecular interactions contribute to the total energy considerably, and the MACE model with $3$~\AA\  cutoff in each layer cannot describe them well. In contrast, the two longer-ranged MACE models have significantly lower energy errors. It is not a coincidence that most short range ML force field models in the literature have cutoffs between 5 and 10~\AA. 

\subsection{Effect of locality beyond RMSE - the dynamics of the bucky-ball catcher}

\begin{figure*}[]
    \centering
    \includegraphics[width=0.9\linewidth]{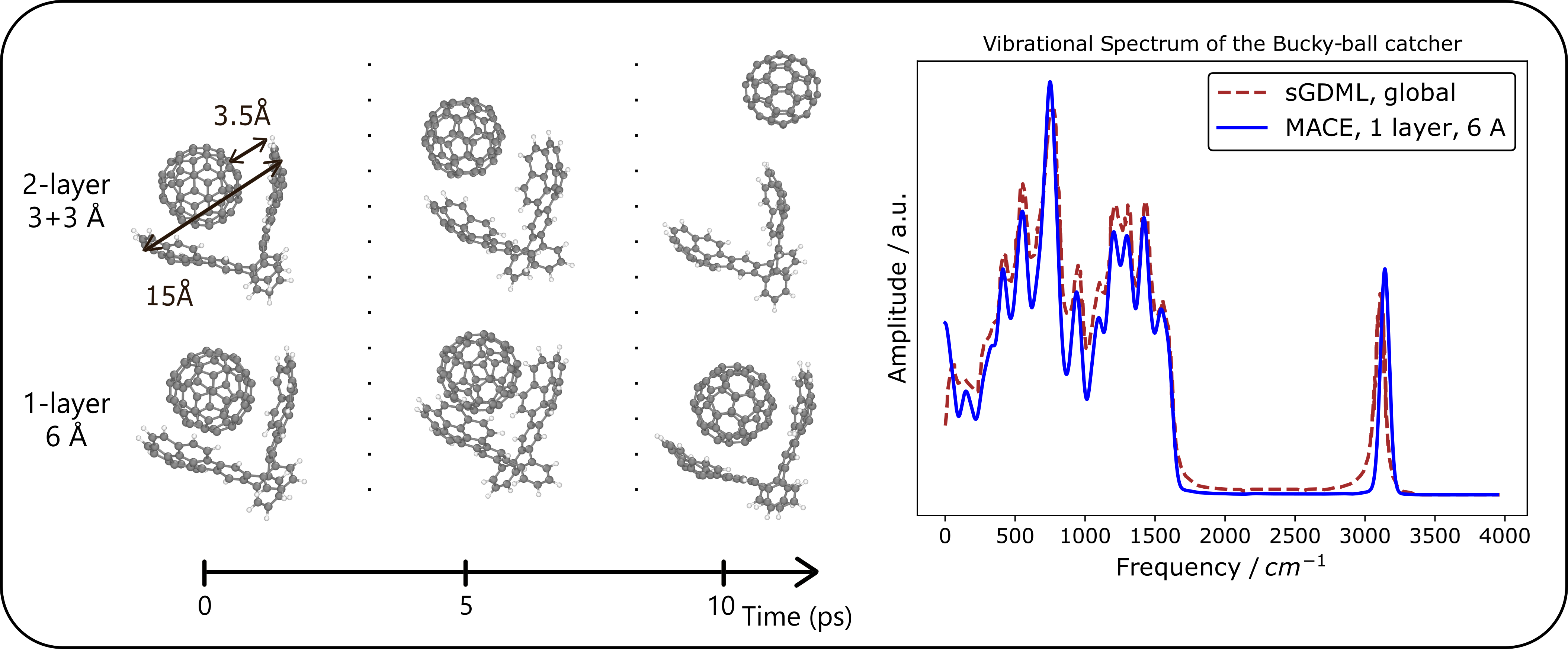}
    \caption{\textbf{Bucky-ball catcher MD} The left panel illustrates that the Bucky-ball catcher dissociates for short cutoff MACE model, but stays together with longer cutoff. The right panel compares the molecular vibrational spectrum computed using local MACE and the global sGDML model }
    \label{fig:bucky_vib}
\end{figure*}

In the following, we investigate the Bucky-ball catcher system in more detail because intermolecular interactions play a crucial role here in the dynamics. Following the experiment in Ref.~\cite{MD22} we ran 200 ps of NVT molecular dynamics simulation with each of the three MACE models introduced in the previous section. As shown on the left panel of Figure~\ref{fig:bucky_vib} for the $2 \times 3$~\AA\ MACE model, the system dissociated into the Bucky-ball and the catcher within 10~ps. The model, which on the whole has excellent accuracy (just 0.5~meV/atom for energies and 13~meV/\AA\ for forces), is unable to fit the attractive dispersion between the two sub-parts of the system because they are typically further than 3~\AA\ apart. On the other hand, the two slightly longer ranged MACE models provide qualitatively correct dynamics. We have also computed the molecular vibrational spectrum obtained as the Fourier transform of the velocity-velocity auto-correlation function to analyse the dynamics quantitatively. On the right panel of Figure~\ref{fig:bucky_vib}, we compare the result of the 1-layer~$6$~\AA\ MACE model to the spectrum published in Ref.~\cite{MD22} computed using sGDML and find excellent agreement including for low frequencies, demonstrating that local models can simulate the dynamics of systems even when intermolecular interactions are important. 

\section{COMP6 - H, C, N, O organic force field}
\label{sec:MACE_ANI}
In this section, we demonstrate that the MACE architecture can be used to train a highly accurate transferable general organic force field for the H, C, N, and O chemical elements. To train the model, we used the subset of the ANI-1x dataset, which contains coupled cluster calculations~\cite{smith2020ani}. We trained a series of MACE models, going from a small (\texttt{64-0}) MACE model to a medium (\texttt{96-1}) and a large (\texttt{192-2}) model. First, we trained the models using the DFT energies and forces and tested them on the COMP6 benchmark suite~\cite{smith2018less}. The result are summarised in Table~\ref{tab:comp6}. We can see that even the smallest MACE model outperforms most previously published models~\cite{smith2018less, zaverkin2023transfer, haghighatlari2022newtonnet, simeon2023tensornet}. The large MACE model improves on the previous state of the art by about a factor of 5, achieving an overall error well below 0.5~kcal/mol. The ANI-MD subset is one where the MACE total energy errors are relatively high compared with the other subsets. This subset is comprised of configurations of 14 different molecules sampled from ANI-MD trajectories. The MACE errors are relatively low on 12 of the 14 molecules. However, on the two largest ones (Chignolin - 149 atoms and TrpCage - 312 atoms), the MACE energy is shifted by a constant value in comparison to the DFT reference energy. The energy-energy correlation plot for all molecules of this subset is shown in the Appendix Figure~\ref{fig:ani_mace_shift}.  The task is particularly challenging because the training set contains very few molecules with more than 50 atoms hence when testing on larger systems there can be an accumulation of small errors that is not controlled in the training. A simple test of this hypothesis could be the addition of a small number of larger molecules to the training set.

We also performed transfer learning to the coupled cluster level of theory using energies only, following Ref.~\cite{zaverkin2023transfer}. The resulting six pre-trained MACE models are published and available to download from GitHub and can be evaluated using the Atomic Simulation Environment (ASE), OpenMM and LAMMPS. 

\begin{table*}[t!]
\label{tab:comp6}
\centering
  \caption{
    \textbf{Mean Absolute Errors on the COMP6 benchmark dataset} Total energies are given in kcal/mol, forces in kcal/mol/ \AA. Note, that the ANI-1x model was trained on $10\times$ more data than the other models.  \textsuperscript{*}For NewtonNet, the decomposition of errors for the subsets was not published and conformations of molecules whose energies were outside a 100~kcal/mol energy range were omitted from the testing. 
  }
     {%
\begin{tabular}{m{2.7cm} c c c c c c c c} 
   \toprule
             &              & \textbf{ANI-1x}  & \textbf{GM-NN}   & \textbf{NewtonNet} & \textbf{TensorNet} & \textbf{MACE}  & \textbf{MACE} & \textbf{MACE}   \\
       &      &      &     &     &      & \texttt{64-0} & \texttt{96-1} & \texttt{192-2} \\
   \midrule
     \multirow{2}{*}{\textbf{ANI-MD}}
        & E  &  3.40   &  3.83  &  -   & \textbf{1.61} &  10.3  &  2.81   &  3.25   \\ 
        & F  &  2.68   &  1.43  &  -   & 0.82 & 1.92  &  0.89   &   \textbf{0.62}  \\
        \hline
     \multirow{2}{*}{\textbf{DrugBank}}
        & E  &  2.65  &  2.78  &  -   & 0.98 & 1.81   &  1.04   &  \textbf{0.73}   \\ 
        & F  &  2.86  &  1.69  &  -   & 0.75 & 1.20   &  0.70   &  \textbf{0.47}   \\
        \hline
     \multirow{2}{*}{\textbf{GDB 7 - 9}}
        & E  &  1.04  &  1.22  &   -  & 0.32 & 0.77  &  0.40   &   \textbf{0.21}  \\ 
        & F  &  2.43  &  1.41  &   -  & 0.53 & 0.96  &  0.54   &   \textbf{0.34}  \\
        \hline
     \multirow{2}{*}{\textbf{GDB 10 - 13}}
        & E  &  2.30  &  2.29  &  -  & 0.83 & 1.54  &  0.88   &   \textbf{0.53}  \\ 
        & F  &  3.67  &  2.25  &  -  & 0.97 & 1.52  &  0.92   &   \textbf{0.62}  \\
        \hline
     \multirow{2}{*}{\textbf{S66x8}}
        & E  &  2.06  &  2.95  &   -  & 0.62 & 1.17  &  0.69   &   \textbf{0.39}  \\ 
        & F  &  1.60  &  0.93  &   -  & 0.33 & 0.65  &  0.33   &   \textbf{0.22}  \\
        \hline
     \multirow{2}{*}{\textbf{Tripeptides}}
        & E  &  2.92  &  3.06  &  -  & 0.92 & 2.10  &  1.18   & \textbf{0.79}   \\ 
        & F  &  2.49  &  1.48  &  -  & 0.62 & 1.09  &  0.66   &  \textbf{0.44}   \\
        \hline
     \multirow{2}{*}{\textbf{COMP6 total}}
        & E  &  1.93  &  2.03  &  1.45\textsuperscript{*} & - & 1.47  &  0.76   &  \textbf{0.48}   \\ 
        & F  &  3.09  &  1.85  &  1.79\textsuperscript{*} & - & 1.31  &  0.77   &  \textbf{0.52}   \\
   \bottomrule
 \end{tabular}
 
}
\end{table*}

\subsection{Biaryl torsion benchmark}
We now evaluate the coupled cluster transfer learned versions of the above MACE organic force fields on the challenging dihedral torsion dataset introduced in Ref.~\cite{lahey2020biaryl_torsions}. This dataset consists of 88 small drug-like molecules with different biaryl dihedral torsional profiles. Such a benchmark is of particular interest in connection with small molecule drug discovery. Accurately describing the torsional barriers is a typical task where classical empirical force fields cannot provide sufficiently accurate descriptions of the potential energy surface~\cite{riniker2018fixed, horton2019qubekit}. 

\begin{figure}
    \centering
    \includegraphics[width=0.9\linewidth]{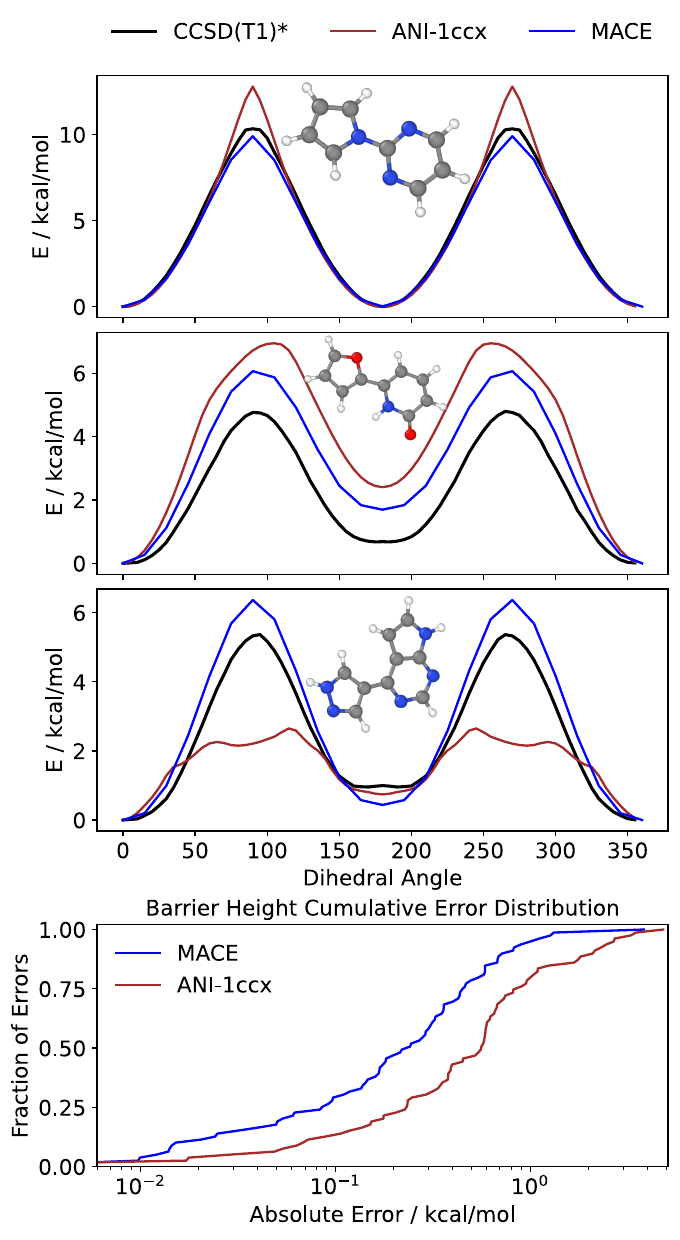}
    \caption{\textbf{Dihedral scans} The top three panels illustrates a selection of challenging dihedral torsional scans computed using MACE \texttt{192-2} from Ref.~\cite{lahey2020biaryl_torsions} where ANI fails. The bottom panel compares MACE \texttt{192-2} and ANI-1ccx torsional barrier height errors. 
    }
    \label{fig:torsions}
\end{figure}

We have evaluated the MACE dihedral torsional scans on the subset of the full dataset which contain only H, C, N and O chemical elements (78 molecules). In the following we evaluate the largest MACE model (\texttt{192-2}) transfer learned to CC level of theory and compare it to ANI-1ccx. We find that the MACE model achieves a mean absolute barrier height error compared to CCSD(T) of 0.36~kcal/mol, improving significantly on the ANI-1ccx model which was identified as the best to date and has an error of 0.78~kcal/mol. We have also compared the cumulative error distribution of MACE and ANI and show it on the bottom panel of Figure~\ref{fig:torsions}. It shows that MACE has very few molecules for which it makes an error close to or larger than 1~kcal/mol. Notably, even the medium and small MACE models do better or close to the same as the ANI model with average barrier height errors of 0.56 and 0.83 kcal / mol. 

All 78 torsional profiles computed with all three CC transfer learned MACE models can be found as part of the Supplementary Information. On Figure~\ref{fig:torsions} we show three examples which were identified as being particularly challenging in the original paper and compare MACE \texttt{192-2} to the ANI-1ccx model. The MACE model produces a smooth surface with the correct position of the minima and maxima in all three cases, similarly to the rest of the test molecules that can be found in the SI.

\section{Vibrational spectrum from 50 coupled cluster calculations}
\label{sec:low_data}

A final test of MACE on small molecular systems is to evaluate its performance in the low-data regime. Data efficiency is crucial in atomistic machine learning as it can save time and computational cost, or enable the models to be trained with more accurate reference data. Recent work has found that the the majority of machine learning potentials are not able to run stable molecular dynamics simulations without iterative fitting even when they are trained on thousands of configuration of a system~\cite{fu2023forces}. Furthermore, the authors found that very low energy and force RMSE did not correlate with the ability to perform stable simulations.  

\begin{figure}
    \centering
    \includegraphics[width=0.85\linewidth]{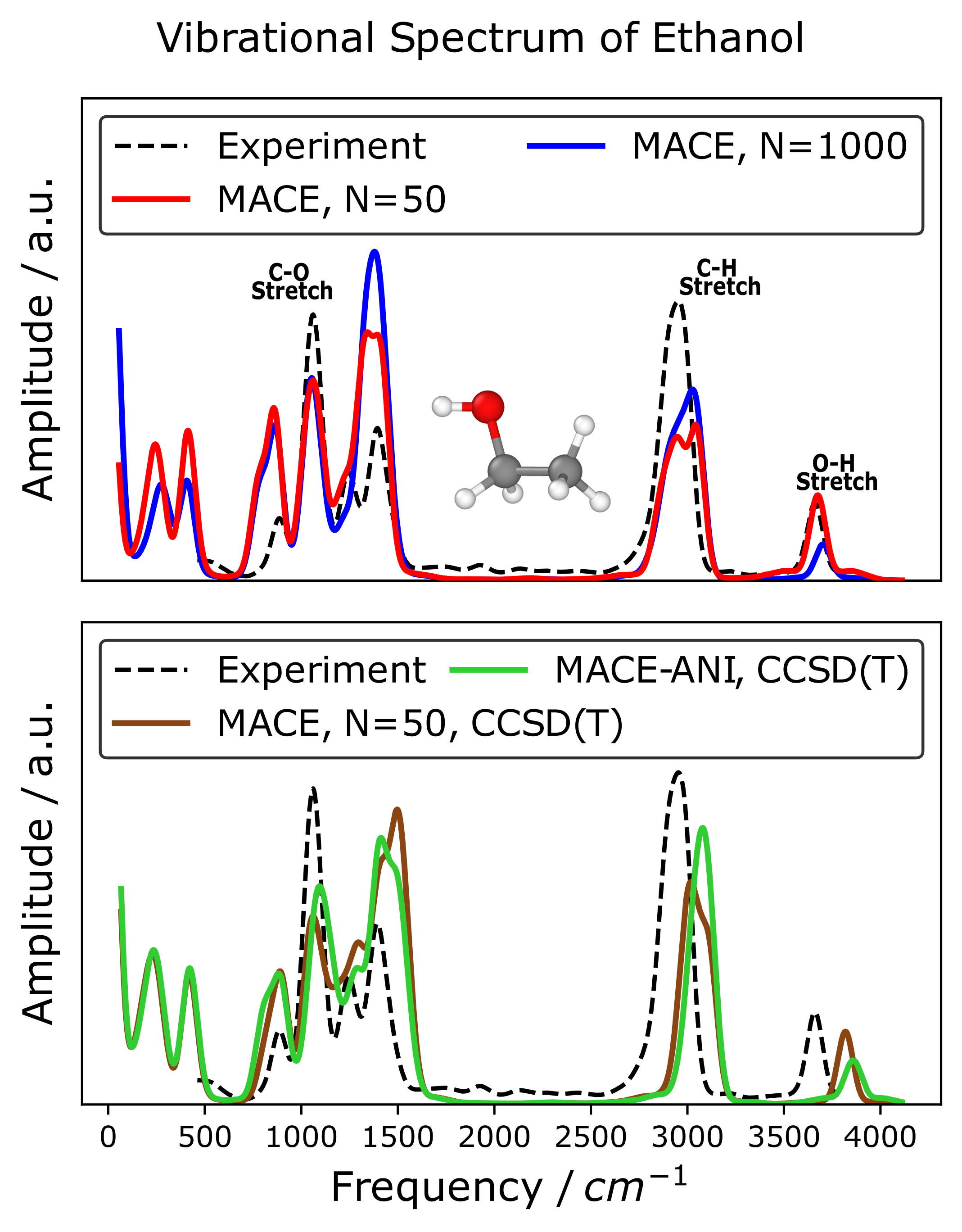}
    \caption{\textbf{Vibrational spectrum of ethanol} The figure illustrates molecular vibrational spectrum computed from the velocity-autocorrelation function. The top panel compares the MACE model fitted to 50 (red) and 1,000 (blue) random revMD17 ethanol geometries. The bottom panel compares the spectrum of the small MACE-COMP6-CC model from Section~\ref{sec:MACE_ANI} (brown) and a MACE model fitted to the same 50 random revMD17 geometries (green) recomputed using CCSD(T) level of theory.
    }
    \label{fig:eth_ir}
\end{figure}

To demonstrate the excellent data efficiency and extrapolation capabilities of MACE we use the \texttt{256-2} MACE models published in Ref.~\cite{batatia2022mace} trained using rMD17~\cite{christensen2020role} and compare the models trained on 50 and 1000 small molecule geometries sampled randomly from the dataset. It was previously shown that 50 configurations are sufficient to train a model that has sub-kcal/mol total energy error.  For three selected systems, ethanol, paracetamol and salicylic acid we ran 50 ps long NVT molecular dynamics simulations which were stable for both training set sizes, obviating the need for any further active learning or iterative fitting.  To validate the dynamics quantitatively we compared the molecular vibrational spectrum. The top panel of Figure~\ref{fig:eth_ir} shows the spectra for ethanol, the two MACE models are in excellent agreement, and are able to reproduce the peaks from the experimental gas-phase IR spectrum of ethanol~\cite{ethanol_IR_spectra}. In the Appendix we also show that the same result holds for paracetamol and salicylic acid. 

With model architectures able to learn an accurate representation of the potential energy surface from so few configurations we can create better models by training to reference data obtained using much more expensive correlated wave-function based quantum mechanical calculations. To demonstrate this we recomputed the 50 training configurations from rMD17 using CCSD(T) level of theory, including the forces~\cite{neese2020orca}. We fitted a MACE model on this data directly, without transfer learning. We also compared the model to the smallest (\texttt{64-0}) general organic MACE model that was transfer learned to coupled-cluster data from Section~\ref{sec:MACE_ANI}. We find that the custom trained and general MACE models are  predicting very similar spectra as shown on the bottom panel of Figure~\ref{fig:eth_ir}. Interestingly, the peaks corresponding to stretching models involving Hydrogens are blue-shifted compared to the experimental positions. This is a result of nuclear quantum effects and the experimental spectrum can be recovered by including these effects via techniques treating the nuclei as quantum particles~\cite{chen2023OH_NQE}. Note, how these shifts are smaller for the original rMD17-derived models - an example of cancellation of errors due to missing correlation in DFT and quantum nuclear effects. 

\section{Amorphous carbon}
\label{sec:carbon}

To demonstrate the capabilities of  MACE on complex solid state systems, we train a model on the carbon dataset published in Ref.~\cite{qamar2022carbon_ACE}. This dataset comprises a wide variety of carbon phases, from crystalline diamond and graphite structures to various amorphous phases and small clusters. The complexity comes from the unique bonding behaviour of carbon atoms that can accommodate 1 to 4 neighbours depending on their oxidation and hybridisation states. Whilst the crystalline phases typically only contain one type of hybridisation of carbon, sp\textsuperscript{2} in graphite and sp\textsuperscript{3} in diamond, in amorphous carbon, these occur simultaneously and can even vary dynamically. Amorphous carbon can be regarded as one of the most challenging systems for machine learning force fields due to its highly complex short- and medium-range electronic structure especially in the low density, sp\textsuperscript{2} dominated phases~\cite{deringer2017carbon_GAP1, rowe2020carbon_GAP2}
\begin{figure}
    \centering 
    \includegraphics[width=1.0\linewidth]{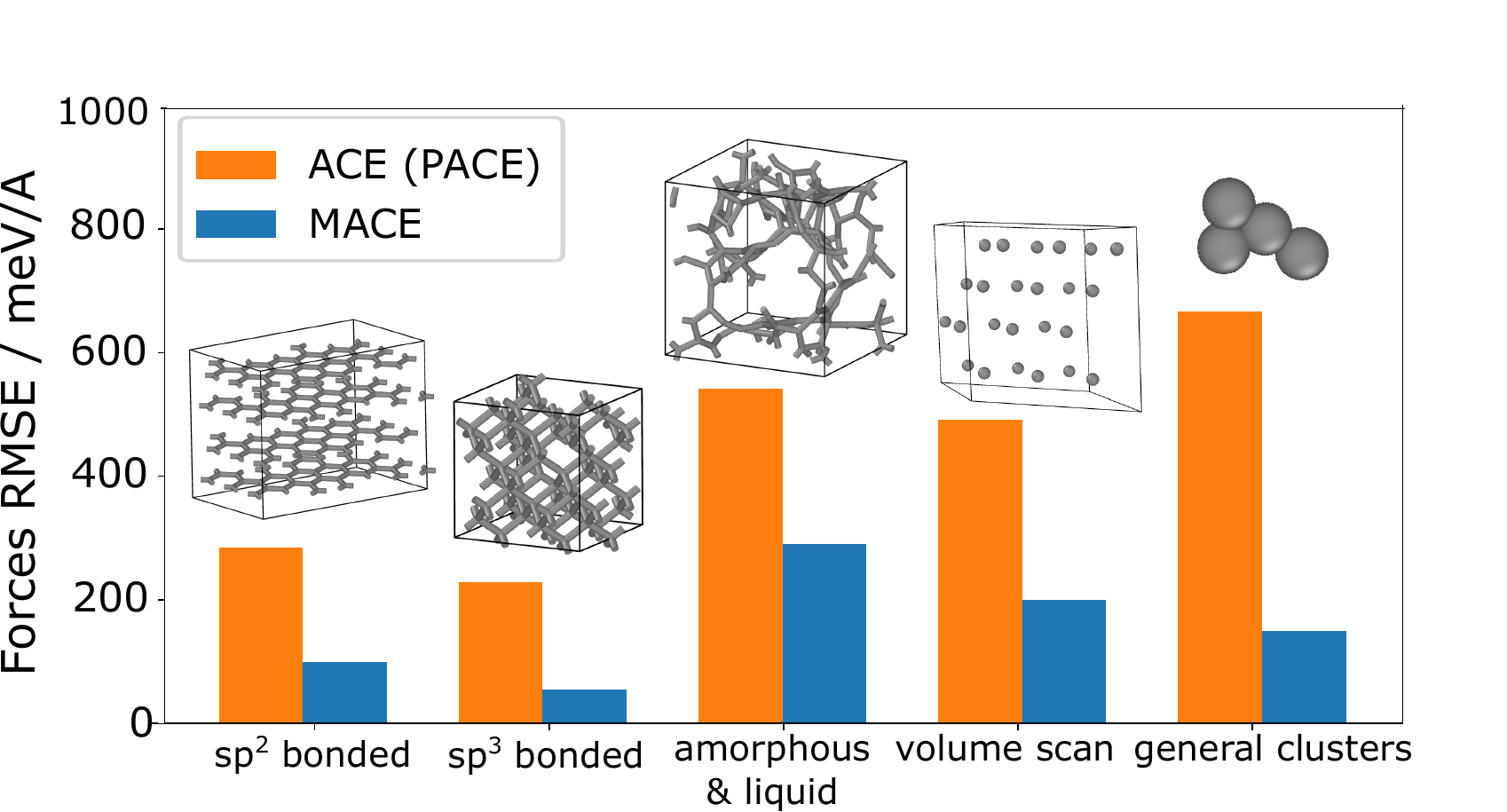}
    \caption{\textbf{Carbon force errors} The figure compares the force errors of the PACE and MACE models for a test dataset consisting of diverse set of 1640 carbon structures~\cite{qamar2022carbon_ACE}. \proof{The PACE errors correspond to the model with equal loss weights for all configurations similarly to the MACE model.}
    }
    \label{fig:carbon_mace}
\end{figure}

Figure~\ref{fig:carbon_mace} compares the test set errors of various subsets of the carbon dataset from Ref.~\cite{qamar2022carbon_ACE}. It shows that MACE \texttt{256-2} significantly improves the state of the art ACE potential across all phases.

\section{HME21 -  37 element disordered crystals}
\label{sec:hme21}
The simultaneous description of a large number of different chemical elements is a long standing challenge for machine learning force fields~\cite{artrith2017BPNN_many_species, willatt2018feature, darby2022TrACE}. The HME21 dataset \cite{takamoto2022HME21, takamoto2022Towards} is made up of a diverse set of disordered and regular crystals from 37 different chemical elements, covering a broad spectrum of chemical space. We train a \texttt{256-2} MACE model on the dataset and compare our results with other previously published models in Table~\ref{tab:hme21}.
\begin{table}[H]
\centering
  \caption{
    \textbf{Energy and Force Mean Absolute Errors on the HME21 dataset}~\cite{takamoto2022HME21}
  }
\resizebox{0.35\textwidth}{!}
     {%
\begin{tabular}{m{3.8cm} c c c}
  \toprule
        &  TeaNet~\cite{takamoto2022teanet}  &  NequIP~\cite{batzner20223NequIP}  &  MACE  \\
        &                                    &                                    & \texttt{256-2} \\
      \midrule
                  Energy (meV/Atom)   &   19.6    &  47.8   & \textbf{16.5} ($\pm 1.2$)   \\
                  Forces (meV/\AA)  &  174   &  199 & \textbf{\textbf{140.2} ($\pm 3.8$)}    \\    
      \bottomrule
  \end{tabular}
}
\label{tab:hme21}
\end{table}
The low errors demonstrate that MACE can effectively handle a wide range of chemical elements and achieves excellent accuracy for both forces and energies. It outperforms the NequIP~\cite{batzner20223NequIP} and TeaNet models~\cite{takamoto2022teanet} by more than 30\% in both metrics.

\revision{
\section{M3GNet: 89 elements condensed materials}

Another challenging condensed phase materials science dataset is the training set of the M3GNet model~\cite{chen2022M3GNet}. This dataset contains an extensive range of materials extracted from the Materials Project~\cite{jain2013MaterialsProject}. The aim of the dataset is to provide a training set for a universal condensed matter interatomic potential of all 89 elements ranging from Hydrogen to Thorium.

\begin{table}[H]
    \centering
    \caption{Energy and Force Mean Absolute Errors on the M3GNet dataset}
    {%
    \begin{tabular}{m{3.5cm} c c }
  \toprule
        &  M3GNet~  &  MACE \\
        & & \texttt{128-2} \\
      \midrule
                  Energy (meV/Atom)   &    34.7   & {\bf 34.1}   \\
                  Forces (meV/\AA)  &  71.7  & {\bf 60.1}   \\  
                  \proof{Stress (GPa)} & {\bf 0.41} &  0.62 \\
      \bottomrule
  \end{tabular}
    }
    \label{tab:my_label}
\end{table}

This dataset is distinct from the previously studied ones due to its greater range of interaction energies and forces. To be able to train the MACE model on such a diverse dataset we used the Huber loss~\cite{huber1992Loss}, as proposed in the M3GNet paper~\cite{chen2022M3GNet}. This loss function switches between an L1 and L2 loss making the learning more tolerant towards outliers. We also found beneficial to slightly generalized the radial basis by making it dependent on the atom's elements and local environment (see \ref{sec:element-radial}).
We observe that an out-of-the-box MACE model with the modified loss and radial function roughly matches the accuracy of M3GNet, improving on forces. There are several possible modifications of MACE that could improve further its performance on training sets with large chemical diversity. These include making the cutoffs element dependent which will be investigated as part of future works. 
}

\section{Liquid Water}
\label{sec:water}

To assess the ability of  MACE to describe complex molecular liquids we fitted a dataset of 1593 liquid water configurations, made up of 64 molecules each~\cite{cheng2019water_bing}. The dataset was computed using the CP2K software~\cite{kuhne2020cp2k} at the revPBE0-D3 level of density functional theory which is known to give a reasonably good description of the structure and dynamics of water at a variety of pressures and temperatures~\cite{marsalek2017water_D}. 

\begin{table}[H]
\centering
  \caption{
    \textbf{Energy and Force errors on liquid water dataset} Ref.~\cite{cheng2019water_bing}
  }
\resizebox{0.5\textwidth}{!}
     {%
\begin{tabular}{m{3.4cm} c c c c c} 
  \toprule
      & \textbf{BP-NN}~\cite{cheng2019water_bing} & \textbf{REANN}~\cite{zhang2021REANN} & \textbf{NequIP}~\cite{batzner20223NequIP} & \textbf{MACE}  & \textbf{MACE}   \\
      &                &        &  (L=2)          &    \texttt{64-0}        & \texttt{192-2} \\
  \midrule
E RMSE (meV / \ce{H2O}) &  7.0  &  2.4  & -   & \textbf{1.9} &       \textbf{1.9}    \\ 
F RMSE (meV / \AA)      &  120  &  53.2 & -   &    37.1      &  \textbf{36.2}   \\ \hline
E MAE (meV / \ce{H2O})  &   -   &    -  & 2.5 &  \textbf{1.2} & \textbf{1.2}    \\ 
F MAE (meV / \AA)       &   -   &    -  & 21  &   20.7        & \textbf{18.5}   \\
  \bottomrule
\end{tabular}

}
\label{tab:water}
\end{table}

\begin{figure*}
    \centering
    \includegraphics[width=0.65\linewidth]{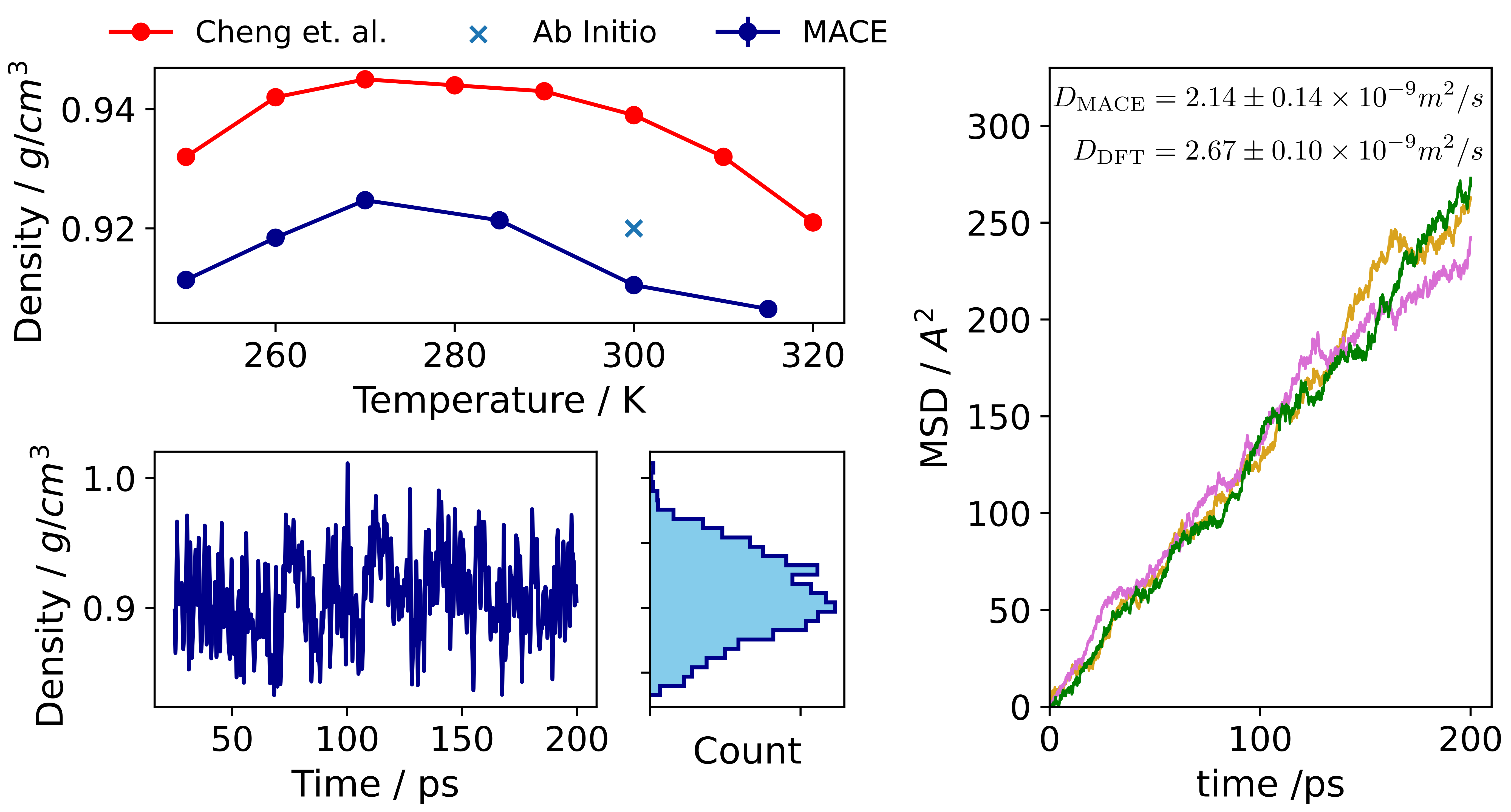}
    \caption{\textbf{Thermodynamic properties of liquid water} The top left panel shows the liquid water density isobar at $p = 1.0$ bar pressure and compares the isobar from from Cheng et. al.~\cite{cheng2019water_bing} and the {\em ab initio} density~\cite{ohto2019water_rho} to MACE \texttt{64-0}. The bottom panel shows an example of the density distribution in the NPT simulations. The right panel shows the mean squared displacement of the water molecules in 3 independent NVT simulations at the equilibrium density (0.91 g / cm\textsuperscript{3}) of MACE. The corresponding DFT value was obtained from a simulation at the DFT equilibrium density of 0.92 g / cm\textsuperscript{3}. 
    }
    \label{fig:water}
\end{figure*}

Table~\ref{tab:water} compares the energy and force errors of the MACE model with other machine learning force fields trained on the same dataset, but using different train test splits. The 3-body Atom-Centred Symmetry Function based feed-forward neural network model (BPNN) has the highest errors. The 3-body invariant message passing model REANN~\cite{zhang2021REANN} and the 2-body equivariant message passing model NequIP~\cite{batzner20223NequIP} significantly improve on the errors of the BPNN model. A further improvement is achieved by the many-body equivariant MACE model. Interestingly a relatively small MACE model using invariant messages, but having an overall body-order of 13 achieves already lower errors than the other best models and the larger MACE model only slightly improves on the force errors.

\subsection{Thermodynamics and kinetics of liquid water}
 To characterise the smaller and faster MACE water model we ran 200 ps  NPT simulations at a range of temperatures from 250 K to 315 K using a combined Nose-Hoover and Parrinello-Rahman barostat~\cite{melchionna1993NPT1, melchionna2000NPT2}. As shown on Figure~\ref{fig:water} the average density was computed after an initial equilibration period at each temperature. The top left panel shows the water density isobar, showing the characteristic density maximum at around 270 K. This is somewhat lower than the experimental value, but is consistent with previous DFT based studies of water~\cite{cheng2019water_bing}. At 300 K we also compare the density of the MACE model with the available {\em ab initio} value from Ref.~\cite{ohto2019water_rho} and find very good agreement. 

Finally, we also investigated a dynamic property, the diffusivity of water which is notoriously difficult to compute accurately. To obtain the diffusivity we took a water configuration from the NPT simulation which had the equilibrium density of 0.91 g / cm\textsuperscript{3} and ran 3 independent 200 ps long NVT simulations. The mean squared displacements from these simulations are shown on the right panel of Figure~\ref{fig:water}. By fitting a linear function on the diffusive part of the MSD we obtained a value of the diffusion coefficient of $2.14\pm 0.14 \times 10^{-9} m^2 / s$ which is in reasonably good agreement with the {\em ab initio} value estimated from much smaller simulations in Ref.~\cite{marsalek2017water_D}. 

\section{The QM9 benchmark}
\label{sec:QM9}
We now test the MACE architecture on tasks that are not directly related to force field fitting  and use the well established QM9 dataset~\cite{ramakrishnan2014QM9}. This benchmark is one of the oldest used to validate machine learning models for chemistry in general, and most of the published machine learning architectures have reported their results on it. Table~\ref{tab:QM9_table} collects the results of these models on 12 different tasks. Notably, MACE achieves state of the art results on 3 of the 4 energy related tasks ($G$, $H$, $U$, $U_{0}$) and also improves the state of the art on one further not-energy related task. 

\begin{table*}[t!]
\centering
  \caption{
    \textbf{Performance of MACE on QM9}
     Mean absolute error (MAE) of various models on the QM9 dataset demonstrating that MACE improves on the state of the art in several tasks. The bold model has the lowest error, whilst the underline indicates models with error within 10\% of the best model for each task. 
     }
\begin{tabular}{lcccccccccccc}
    \toprule
        & \textbf{Gap} & \textbf{Homo} & \textbf{Lumo} & $\bm{C_{V}}$ & $\bm{\mu}$ & \textbf{ZPVE} & $\bm{R^{2}}$ & $\bm{\alpha}$ & $\bm{G}$ & $\bm{H}$ & $\bm{U}$ & $\bm{U_{0}}$ \\ 
        & meV & meV & meV & cal/mol K & D & meV & $\alpha_0^2$ & $\alpha_0^3$ & meV & meV & meV & meV \\
        \midrule
        \textbf{NMP}~\cite{gilmer2017NMP} & 69 & 43 & 38 & 0.040 & 0.030 & 1.50 & 0.180 & 0.092 & 19 & 17 & 20 & 20 \\ 
        \textbf{SchNet}~\cite{schutt2017schnet} & 63 & 41 & 34 & 0.033 & 0.033 & 1.70 & 0.073 & 0.235 & 14 & 14 & 19 & 14 \\ 
        \textbf{Cormorant}~\cite{anderson2019cormorant} & 61 & 34 & 38 & 0.026 & 0.038 & 2.03 & 0.961 & 0.085 & 20 & 21 & 21 & 22 \\ 
        \textbf{LieConv}~\cite{finzi2020LieConv} & 49 & 30 & 25 & 0.038 & 0.032 & 2.28 & 0.800 & 0.084 & 22 & 24 & 19 & 19 \\ 
        \textbf{DimeNet++}~\cite{gasteiger2020DimeNet++} & 33 & 25 & 20 & \underline{0.023} & 0.030 & 1.21 & 0.331 & 0.044 & 7.6 & 6.5 & 6.3 & 6.3 \\ 
        \textbf{EGNN}~\cite{satorras2021EGNN} & 48 & 29 & 25 & 0.031 & 0.029 & 1.55 & 0.106 & 0.071 & 12 & 12 & 12 & 11 \\ 
        \textbf{PaiNN}~\cite{schutt2021PAINN} & 46 & 28 & 20 & 0.024 & \underline{0.012} & 1.28 & 0.066 & 0.045 & 7.4 & 6.0 & 5.8 & 5.9 \\ 
        \textbf{TorchMD-NET}~\cite{tholke2022TorchMDNet} & 36 & 20 & 18 & 0.026 & \textbf{0.011} & 1.84 & \textbf{0.033} & 0.059 & 7.6 & 6.2 & 6.4 & 6.2 \\ 
        \textbf{SphereNet}~\cite{liu2022SphereNet} & 32 & 23 & 18 & \underline{0.022} & 0.026 & \underline{1.12} & 0.292 & 0.046 & 7.8 & 6.3 & 6.4 & 6.3 \\ 
        \textbf{SEGNN}~\cite{brandstetter2022SEGNN} & 42 & 24 & 21 & 0.031 & 0.023 & 1.62 & 0.660 & 0.060 & 15 & 16 & 13 & 15 \\ 
        \textbf{EQGAT}~\cite{le2022EQGAT} & 32 & 20 & 16 & 0.024 & \textbf{0.011} & 2.00 & 0.382 & 0.053 & 23 & 24 & 25 & 25 \\ 
        \textbf{Equiformer}~\cite{liao2023equiformer} & 30 & \textbf{15} & \textbf{14} & \underline{0.023} & \textbf{0.011} & 1.26 & 0.251 & 0.046 & 7.6 & 6.6 & 6.7 & 6.6 \\ 
        \textbf{MGCN}~\cite{lu2019MGCN} & 64 & 42 & 57 & 0.038 & 0.056 & \underline{1.12} & 0.110 & \textbf{0.030} & 15 & 16 & 14 & 13 \\
        \textbf{Allegro}~\cite{musaelian2023Allegro} & - & - & - & - & - & - & - & - & \underline{5.7} & \underline{4.4} & \underline{4.4} & 4.7 \\ 
        \textbf{NoisyNodes}~\cite{godwin2021NoisyNodes} & 29 & 20 & 19 & 0.025 & 0.025 & \underline{1.16} & 0.700 & 0.052 & 8.3 & 7.4 & 7.6 & 7.3 \\ 
        \textbf{GNS-TAT+NN}~\cite{zaidi2023pretrainingNN} & \textbf{26} & \underline{17} & 17 & \underline{0.022} & 0.021 & \textbf{1.08} & 0.65 & 0.047 & 7.4 & 6.4 & 6.4 & 6.4 \\
        \textbf{Wigner Kernels}~\cite{bigi2023wignerKernels} & - & - & - & - & - & - & - & - & - & - & - & \underline{4.3 } \\
        \textbf{TensorNet}~\cite{simeon2023tensornet}  & - & - & - & - & - & - & - & - & \underline{6.0} & \textbf{4.3 } & \underline{4.3 } & \underline{4.3 } \\
        \hline 
        \\
        \textbf{MACE} & 42 & 22 & 19 & \textbf{0.021} & 0.015 & 1.23 & 0.210 & 0.038 & \textbf{5.5} & \underline{4.7} & \textbf{4.1} & \textbf{4.1} \\[3pt]
    \bottomrule
    \end{tabular}
  \label{tab:QM9_table}
\end{table*}

To achieve the best result on intensive properties, such as $C_V$, we  modified the readout function of the MACE model. This was necessary because the simple sum pooling operation of the node outputs that is used in the MACE force fields, results in size-extensive overall prediction. The precise form of the intensive global non-linear readout function used is described in the SI. 

\begin{figure}
    \centering
    \includegraphics[width=1.0\linewidth]{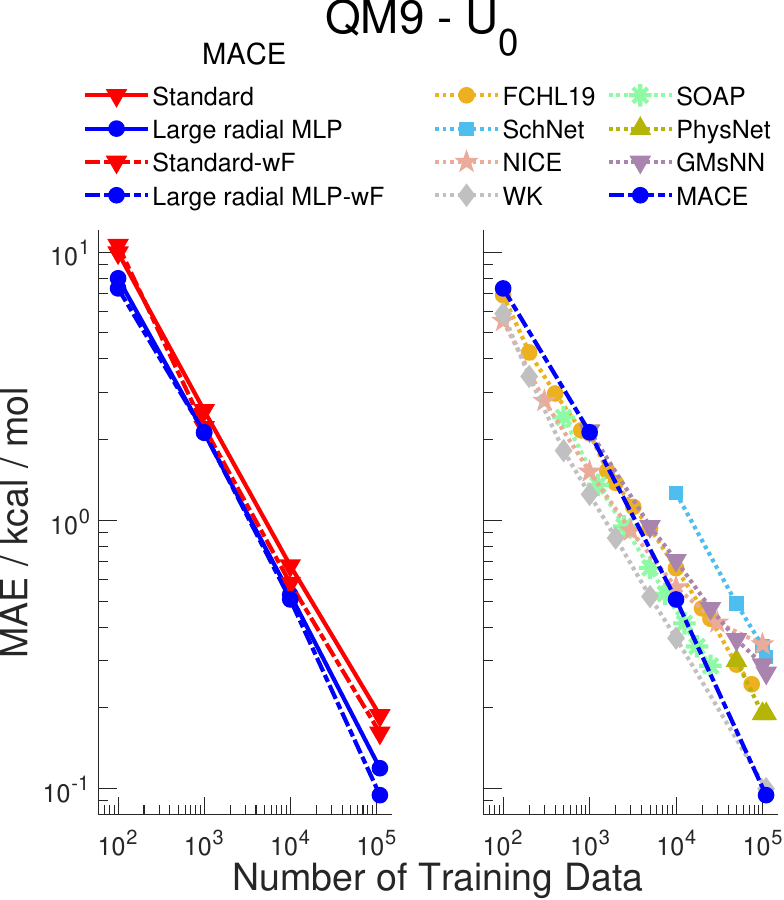}
    \caption{\textbf{QM9 Learning Curves} The left panel compares the standard large MACE model (256-2) to a modified version for QM9 where the size of the MLP in the radial basis was increased. The figure also shows that including the 0 force information (labelled -wF) in the training leads to lower errors. The right panel compares the best MACE with a number of other machine learning models)
    }
    \label{fig:qm9}
\end{figure}

\paragraph{Learning curve on Potential Energies}
Finally, it is interesting to examine the potential energy (U\textsubscript{0}) models in more detail. On the left panel of Figure~\ref{fig:qm9}, the learning curve of 4 different MACE models are compared. The blue models use hyperparameters of MACE that are similar to the ones used throughout the paper (\texttt{256-2}). In comparison, the models denoted with red are a significantly larger versions of MACE, which use a much larger MLP in the learnable radial basis (see Eq.~\eqref{eq:radial_MLP}) which has been shown to help for QM9 other models like Allegro~\cite{musaelian2023Allegro}. This extra flexibility results in higher accuracy, especially in the very high data regime. A further interesting point is the comparison of the models trained using energies only (the usual practice for QM9, denoted with the solid line) with the models whose loss function also included forces exploiting the knowledge that QM9 is made up of equilibrium geometries with zero forces on all the atoms. This extra information (but needing no new QM calculations) increases the accuracy of both the small and large MACE models. 

In the right panel of Figure~\ref{fig:qm9}, we compare the learning curve of the best MACE model to the learning curve of several of the other very good models. It shows that kernel models such as FCHL~\cite{christensen2020fchl}, SOAP~\cite{bartok2013SOAP}, the linear NICE model~\cite{nigam2020NICE}, the feed forward neural network GMsNN~\cite{zaverkin2021GMNN} and the message passing SchNet~\cite{schutt2017schnet} and PhysNet~\cite{unke2019physnet} models all achieves comparable errors. The models that can surpass this significantly are Allegro~\cite{musaelian2023Allegro}, Wigner kernels~\cite{bigi2023wignerKernels} and MACE, all of which use higher body-order features. This strongly hints that high body-order is a crucial property of the most successful atomistic ML models. 

\section{Conclusion}
\label{sec:conclusions}
In this paper we have demonstrated the power of the MACE architecture on a wide variety of chemical tasks and systems. We have investigated the effect of the locality assumption on large molecular systems and demonstrated that a two layer MACE model with a 5 \AA\ cutoff on each layer is able to improve on the state of the art global model in accuracy by up to a factor of 10. We also showed that the molecular vibrational spectrum can be accurately reproduced using the local MACE model. We have also demonstrated the transferability of the MACE model by fitting to just 10\% of the entire ANI-1x dataset. MACE improved on the state of the art on the COMP6 comprehensive organic molecule benchmark by a factor of 5 and 7 for energies and forces respectively compared to the ANI model and a factor of 3-4 compared to the equivariant message passing NewtonNet model. We showed that the models can be transfer learned to coupled-cluster level of theory and used a CC-transfer learned MACE model to compute the vibrational spectrum of ethanol. Further tests demonstrated excellent data efficiency and extrapolation by computing the coupled cluster level vibrational spectrum of ethanol using a MACE model trained on just 50 reference calculations without the need for iterative training. We have also shown excellent accuracy on condensed phase systems such as diverse phases of carbon and liquid water. Our results on the M3GNet dataset show that MACE reaches excellent accuracy and transferability on large and chemically diverse datasets, paving the way to new state of the art universal force fields for materials.  Finally, we also demonstrated that MACE improves on the state of the art for many properties of the standard QM9 quantum chemistry benchmark. 

Overall, these results set a new standard for machine learning force fields in chemistry and demonstrate that the MACE architecture is capable of describing a wide variety of systems with little to no modification to the model training and hyperparameters. This could ultimately lead to force fields with unprecedented accuracy created at ease without significant user input or requirements of expertise.

\section*{Supplementary Material}
The Supplementary Material contains all torsion drives for the biaryl set molecules. 

\section*{Acknowledgement}
DPK acknowledges support from AstraZeneca and the EPSRC. We used computational resources of the UK HPC service ARCHER2 via the UKCP consortium and funded by EPSRC grant EP/P022596/1. This work was also performed using resources provided by the Cambridge Service for Data Driven Discovery (CSD3). We would like to thank Michele Ceriotti for providing the raw data for the QM9 learning curves. We would also like to thank Joshua Horton for helping with the implementation of MACE into the QCEngine driver for the torsional scans. We would like to thank Mario Geiger for the implementation of MACE in JAX. For the purpose of open access, the authors have applied a Creative Commons Attribution (CC BY) licence to any Author Accepted Manuscript version arising from this submission.

\section*{Conflict of Interest}
The authors have no conflicts to disclose. 

\section*{Data Availability}
The data supporting the findings of the article are included in the manuscript and its Supplementary Information. The MACE code is freely available on \url{https://github.com/ACEsuit/mace} and example input scripts and pre-trained models to reproduce the results are provided in the MACE code documentation website: \url{https://mace-docs.readthedocs.io/en/latest/examples/training_examples.html}. The datasets used are all published elsewhere and referenced in the text. The JAX version of MACE is freely available on \url{https://github.com/ACEsuit/mace-jax}.

\nocite{*}
\bibliography{references}

\appendix

\section*{Appendix}

\renewcommand{\thesubsection}{\Alph{subsection}}
\counterwithin*{equation}{subsection}

\subsection{Effect of the loss scheduler}
\begin{figure}
    \centering
    \includegraphics[width=1.0\linewidth]{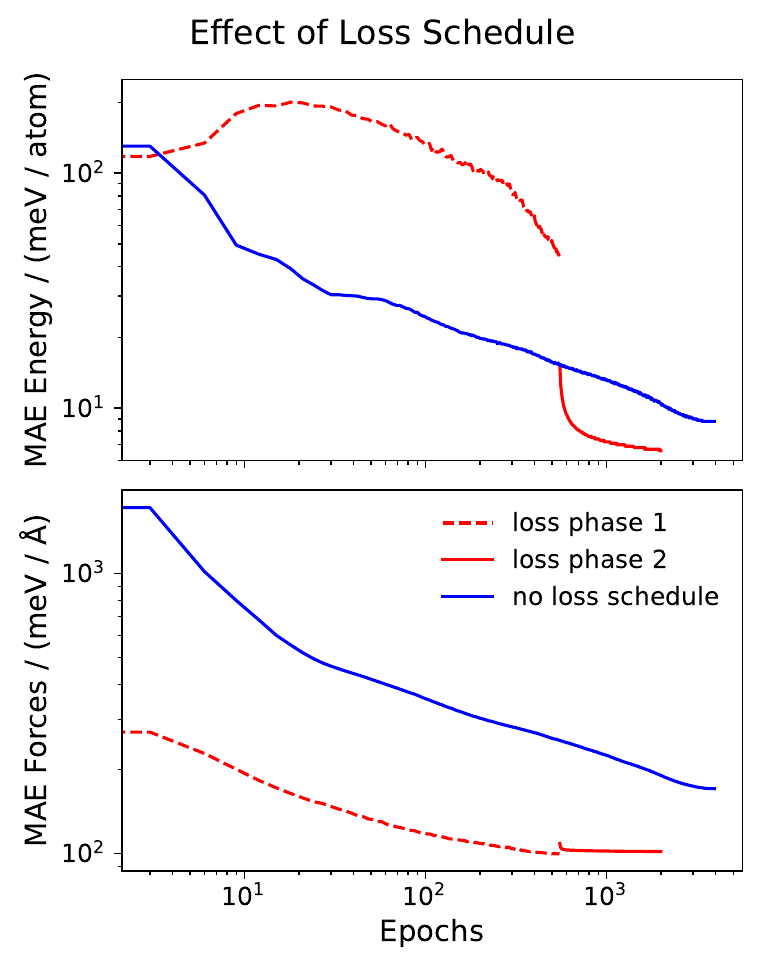}
    \caption{\textbf{Two phases of learning} The energy and force validation errors of the \texttt{64-1} MACE model trained on a subset of Section~\ref{sec:MACE_ANI} training data. The second phase of the loss schedule decreases the energy errors, \revision{and the two-phase learning results in decreased errors compared to using the second phase loss only.}}
    \label{fig:loss_schedule}
\end{figure}

Figure~\ref{fig:loss_schedule} demonstrates how the changing loss weights result in a dramatic improvement of the energy errors for the medium 96-1 MACE model from Section~\ref{sec:MACE_ANI} without significantly affecting the force accuracy. 

\subsection{Old Loss Function}
\label{sec:old_loss}
Most of the models in the paper have been trained using an alternative loss function,
\renewcommand{\theequation}{\thesubsection.\arabic{equation}}
\begingroup\makeatletter\def\f@size{8}\check@mathfonts
\def\maketag@@@#1{\hbox{\m@th\large\normalfont#1}}%
\begin{align}
  \label{eq:loss_fn_old}
  \mathcal{L} = 
  \frac{\lambda_{E}}{B} \sum_{b=1}^{B} &\left(\frac{E_{b} - \hat{E}_{b}}{N_{b}} \right)^{2} + \\ \notag
  &+ \frac{\lambda_{F}}{3B} \sum_{b=1}^{B} 
  \frac{1}{N_{b}} 
  \sum_{i_{b},\alpha=1}^{N_{b}, 3} 
  \left(- \frac{\partial E_{b}}{\partial r_{i_{b},\alpha}} - \hat{F}_{i_{b},\alpha} \right)^{2} ,
\end{align}
\endgroup
This loss is less intuitive as the energy is per atom, but the forces are per atom squared because of the additional averaging. This results in much larger force weights of up to 1,000:1 compared to energy weights to counterbalance this effect. With the loss function in the main text the typical weights have ratio 10:1. All model in this paper, except where otherwise stated, were trained using the loss function in Eq~\eqref{eq:loss_fn_old}.

\subsection{Element dependent radial}
\label{sec:element-radial}

The radial basis of equation~\eqref{eq:phi-basis-t} is agnostic to the element of both the sender and receiver atoms. While we observe this choice performs very well for datasets with few elements (up to 10), it can be beneficial for datasets that include many elements to expend this radial as a function of the element of the atoms.
Therefore, we input the radial MLP and the scalar part of the previous node features of both the sender and receiver atom.
\begingroup\makeatletter\def\f@size{8}\check@mathfonts
\def\maketag@@@#1{\hbox{\m@th\large\normalfont#1}}%
\begin{align}
  \label{eq:dependent_radial}
    &R_{k \eta_{1} l_{1}l_{2} l_{3}}^{(s)}(r_{ij}, \{h_{i, k00}\}_{k}, \{h_{j, k00}\}_{k}) = \notag \\   &{\rm MLP}\left( \left\{ {j_0^n} (r_{ij})\right\}_{n}, \{h_{i, k00}\}_{k}, \{h_{j, k00}\}_{k} \right) \\
      &\phi_{ij,k \eta_{1} l_{3}m_{3}}^{(s)} = 
    \sum_{l_1l_2m_1m_2} C_{\eta_1,l_1m_1l_2m_2}^{l_3m_3} \times \notag\\
      & R_{k \eta_{1} l_{1}l_{2} l_{3}}^{(s)}(r_{ij}, \{h_{i, k00}\}_{k}, \{h_{j, k00}\}_{k}) Y^{m_{1}}_{l_{1}} (\boldsymbol{\hat{r}}_{ij}) \bar{h}^{(s)}_{j,kl_2m_2}
  \end{align}
\endgroup
where $h_{i, k00}, h_{j, k00}$ corresponds to the scalar part of the node features of atom $i$ and $j$.

\subsection{Computational details}

All models were trained using NVIDIA A100 GPU-s. The training time varied between the datasets, for small molecules it was between 1-5 hours, for the larger datasets it depended on the model size amd in the case of the ANI-1x and QM9 datasets and largest models was 3-4 days. We have stopped the training once the validation loss stopped improving. 

All experiments were carried out using the MACE code available at \url{https://github.com/ACEsuit/mace}. Example input files for the experiments can be found in the Examples page of the MACE documentation at: \url{https://mace-docs.readthedocs.io}

\paragraph{Common settings to all MACE Force Field models} All MACE force field models fitted in Section~\ref{sec:locality} to Section~\ref{sec:water} shared most of their core training and model settings. They all used two MACE layers (except where explicitly stated in MD22, where we compared to a single layer model), they all used $l_{max}=3$ in the spherical harmonic expansion, they all used 8 radial Bessel features with polynomial envelope for the cutoff with $p=5$. The radial features well fed into an MLP of size [64, 64, 64] using SiLu non-linearities. All models used correlation order $N=3$ in the MACE layer (4-body messages).  The readout function of the first layer was a simple linear transform, whereas for the second layer it is a single-layer MLP with 16 hidden dimensions. Models were trained with AMSGrad variant of Adam with default parameters $\beta_1 = 0.9$, $\beta_2 = 0.999$, and $\epsilon=1e-8$. We used a learning rate of 0.01 and an on-plateau scheduler that was reducing the learning rate when the validation loss did not improve for a certain number of epochs. We used an exponential moving average with weight 0.99 when updating the weights both for the validation set evaluations and the final models. The loss function used is the one given in Equation~\eqref{eq:loss_fn_old}. 

\paragraph{MD22 models}
The three MACE models fitted to the MD22 dataset all used 256 uncoupled feature channels. The two layer models used $L_{max} = 2$ for the messages. 95\% of the training st was used for training and a random 5\% subset was used for validation. The energy was normalised by shifting by the mean of the potential energies. The initial weights in the loss function were: $\lambda_E = 10$, $\lambda_F = 1000$ followed by a switch to $\lambda_E = 1000$, $\lambda_F = 10$ with the learning rate being reduced to 0.001.  

\paragraph{COMP6 models}
The three models all used $r_{\text{cut}} = 5~$\AA\ cutoff radius. The number of uncoupled feature channels was 64, 96 and 192 with the message equivariance order $L_{max}$ being also increased from 0 (invariant) to 1 and 2 for the three models respectively. The number of parameters for the three models were 196,944, 517,872 and 2,229,456, respectively. We used loss weights: $\lambda_E = 40$, $\lambda_F = 1000$ followed by $\lambda_E = 1000$, $\lambda_F = 10$ in the second phase of the training. For normalisation of the energy we were using the atomization energy, meaning that we applied a fixed shift to each molecule using the isolated atom energies computed with DFT which were published as part of the original dataset. 

\paragraph{revMD17 models}
The models used here were identical to the ones published in \cite{batatia2022mace}. 

\paragraph{Amorphous carbon model} 
The carbon MACE model was trained on the carbon dataset published in~\cite{qamar2022carbon_ACE}. We created a training, validation and test split by selecting $10 \%$ of each configuration type to be in test and randomly selected $10 \%$ for the validation set from the training set. We used a MACE model with two layers, 256 feature channels, $l_{\text{max}} = 3$ spherical harmonics and passing $L_{\text{max}} = 2$ messages. We used a cutoff of $r_{\text{cut}} = 6.0$ \AA. The standard weighted energy forces loss was used, with a weights of  $\lambda_E = 80$, $\lambda_F = 1000$ for the first 225 epochs and $\lambda_E = 1000$, $\lambda_F = 100$ for the 75 last epochs. The  batch size was 5. The weight decay of $5e-7$ was similarly applied to a selected weights as in \cite{batatia2022mace}.

\paragraph{HME21}
The MACE model was trained using the published train-test split. The training data was reshuffled after each epoch. We used 256 uncoupled feature channels, $l_{max}=3$ and invariant messages $L_{max}=2$. We used a cutoff of 6 \AA . The standard weighted energy forces loss was used, with a weight of 1 on energies and a weight of 10 on forces. 
\revision{\paragraph{M3GNet}
The MACE model was trained using the published train-test split. The training data was reshuffled after each epoch. We used 128 uncoupled feature channels, $l_{max}=3$ and equivariant messages $L_{max}=2$. We used a cutoff of 5 \AA . We use the Huber loss with $\delta=0.01$ with loss weights of 1 on forces, 1 on energies and 0.1 on stresses.}
\paragraph{Liquid water model}
The two models both used a cutoff of $r_{\text{cut}} = 6$\AA. The small model used 64 uncoupled channels and invariant messages ($L_{max} = 0$), whereas the larger model used 192 channels and $L_{max} = 2$. The weights used for the training were $\lambda_E = 10$, $\lambda_F = 1000$ followed by $\lambda_E = 1000$, $\lambda_F = 10$ in the second phase of the training. 

\paragraph{QM9 models}
For the $U_0$ property we ran experiments with a standard MACE model identical to the one used for the MD22 tests, meaning 256 uncoupled channels and $L_{max} = 2$. We have found that a slightly modified version of MACE which uses a much increased radial MLP achieves superior error compared to the standard one. This model used a radial MLP of size [128, 256, 512, 1024]. We have also reduced the learning rate to 0.001, increased the width of the readout MLP from 16 to 128 and applied an increased exponential moving average exponent of 0.999. Furthermore, for the case of intensive properties (gap, homo, lumo, $\mu$, C\textsuperscript{V} and zpve) we have changed the readout function from applying a sum pooling which results in a size extensive output, as desired for energy like quantities, into a non-linear pooling operation which first applies both sum, mean and standard deviation pooling, followed by the application of an attention mechanism to form the final output. We found that such a readout function led to up to two fold improvement compared to the simple linear extensive pooling operation.

\subsection{Effect of locality on molecular dynamics simulations}
\paragraph{Tetrapeptide}

\begin{figure}
    \centering
    \includegraphics[width=0.8\linewidth]{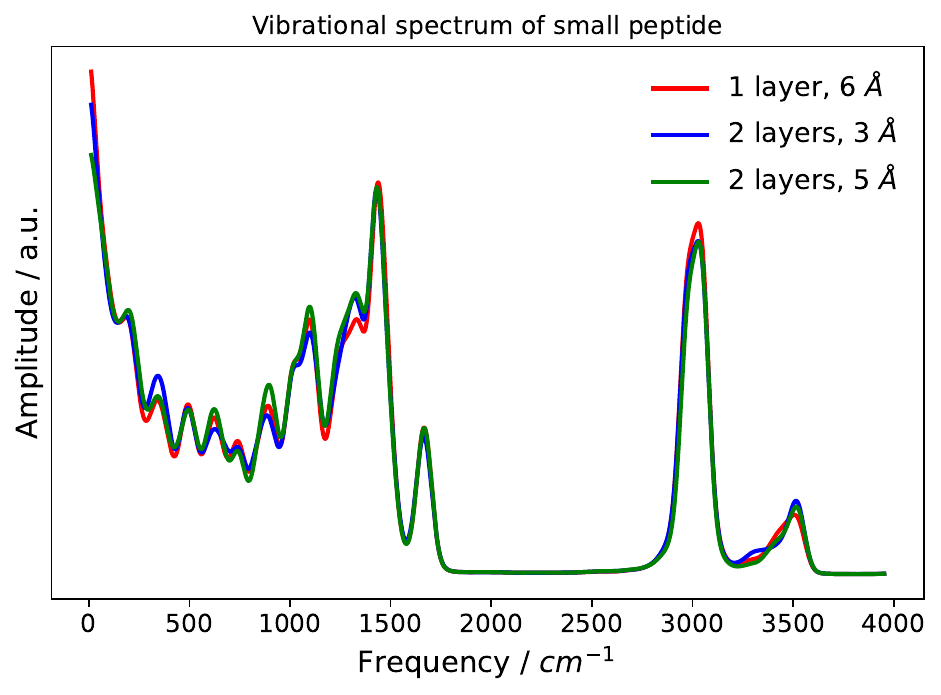}
    \caption{Vibrational spectrum of the tetrapeptide from the MD22 dataset}
    \label{fig:prot_vib}
\end{figure}

On Figure~\ref{fig:prot_vib} we show that the vibrational spectrum of the short range and longer range MACE model agrees remarkably well for the tetrapeptide system of MD22. This demonstrates that the locality assumption can be valid and short-range models can be used to compute accurate vibrational spectrum of even larger systems.

\subsection{COMP6: ANI-MD molecules}

\begin{figure*}
    \centering
    \includegraphics[width=0.53\linewidth]{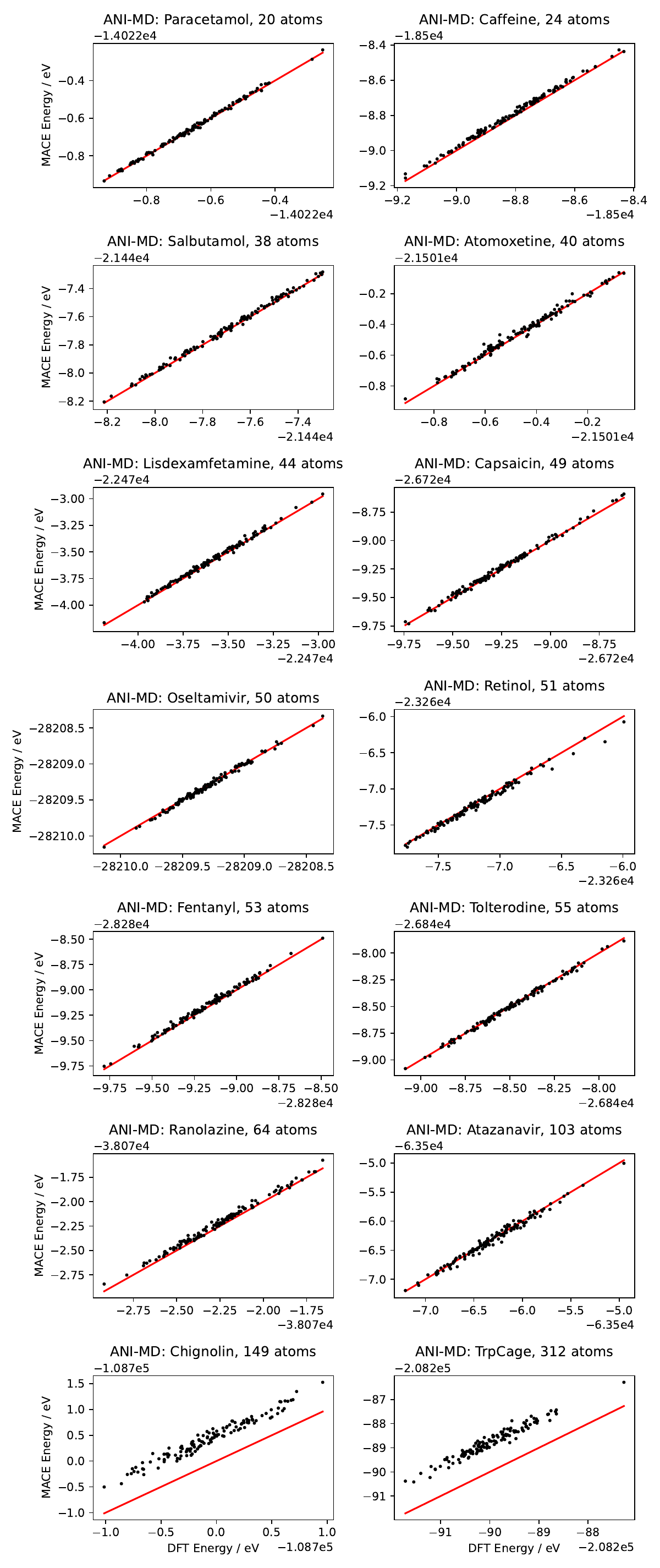}
    \caption{ANI-MD subset of COMP6 showing MACE vs DFT energy correlation plot}
    \label{fig:ani_mace_shift}
\end{figure*}

Figure~\ref{fig:ani_mace_shift} shows that 12 of the 14 molecules have very low errors scattered randomly around the DFT value. The two larger peptides have a systematic shift compared tot he ground truth value. 

\subsection{Vibrational spectrum from 50 QM calculations}

\begin{figure*}
    \centering
    \includegraphics[width=0.8\linewidth]{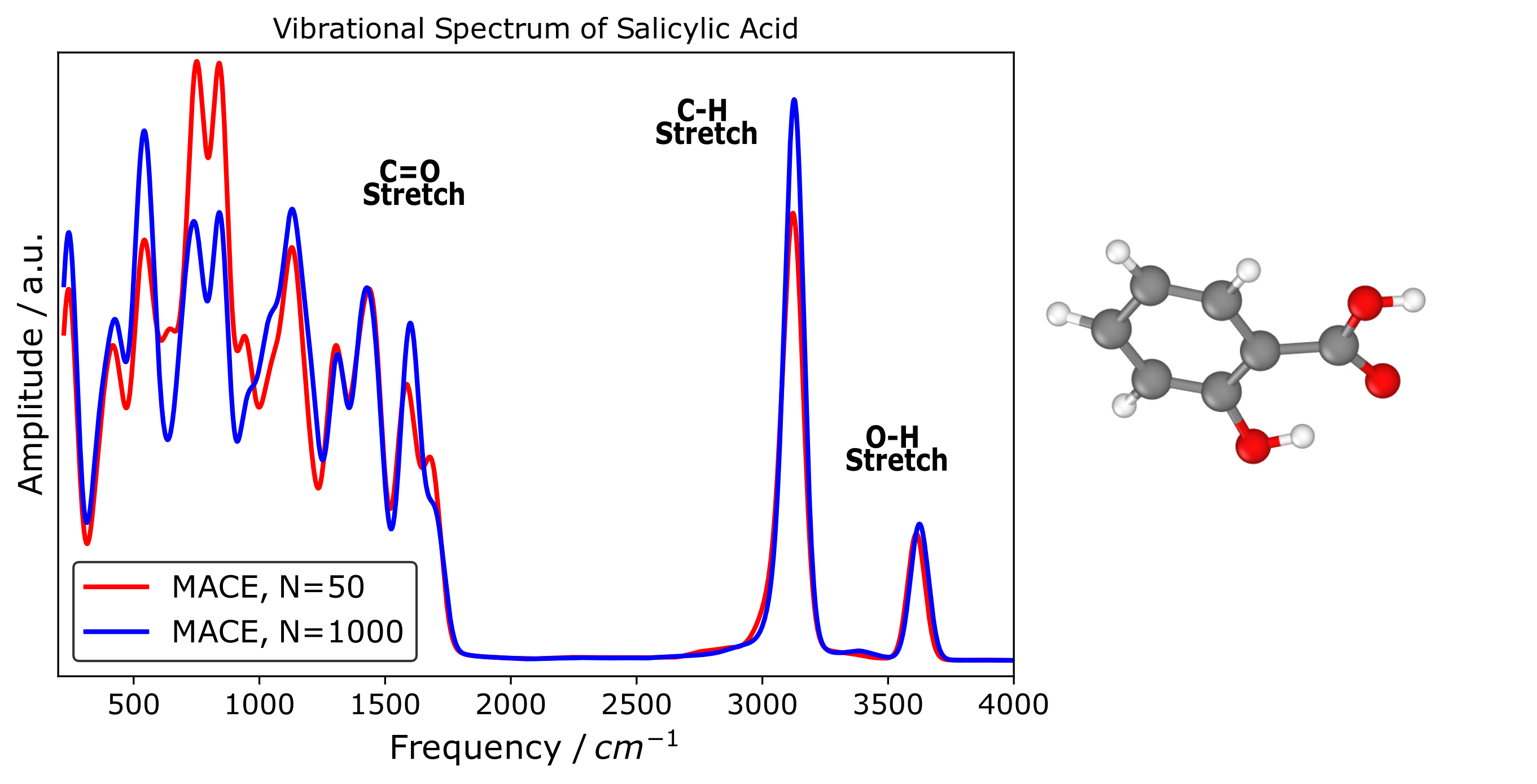}
    \caption{\textbf{Vibrational spectrum of Salicylic acid} 
    }
    \label{fig:sal_ir}
\end{figure*}

\begin{figure*}
    \centering
    \includegraphics[width=0.8\linewidth]{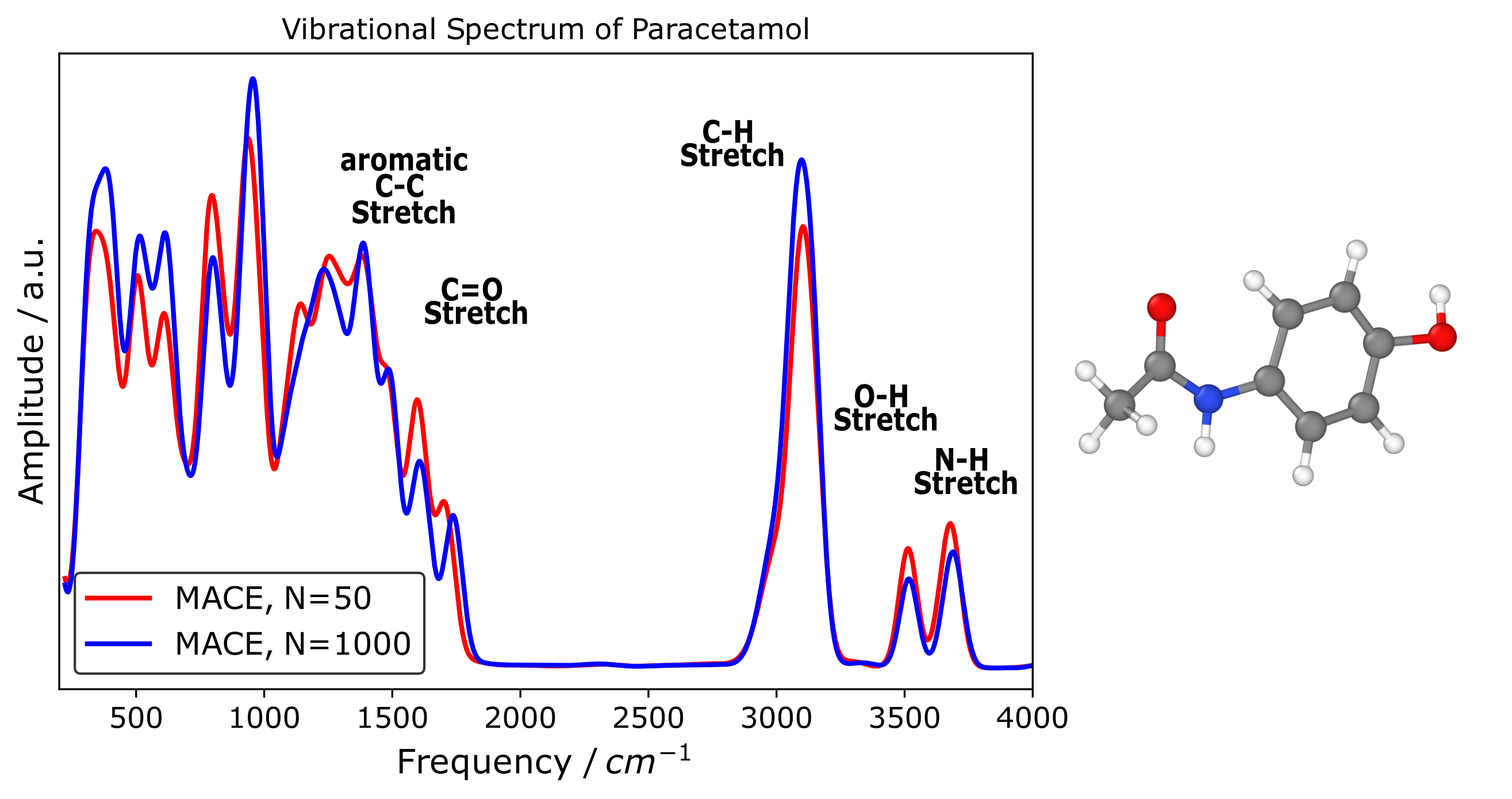}
    \caption{\textbf{Vibrational spectrum of Paracetamol} 
    }
    \label{fig:par_ir}
\end{figure*}

On Figure~\ref{fig:sal_ir} and \ref{fig:par_ir} we show two further examples of accurate MACE molecular vibrational spectrum using a model fitted to just 50 QM calculations. The salicylic acid and paracetamol molecules represent more challenging test cases than ethanol in the main text. 

\end{document}